\documentclass[preprint,12pt]{elsarticle}

\usepackage{graphicx}
\usepackage{float}
\usepackage{amssymb}

\usepackage{amsmath}

\usepackage{multicol}
\usepackage{multirow}
\usepackage{tabularx}
\usepackage{color,soul}

\usepackage{comment}
\usepackage{natbib}
\usepackage{mathtools}
\usepackage{caption}
\usepackage{subcaption}
\usepackage{bibunits}
\usepackage[colorinlistoftodos]{todonotes}
\usepackage{framed}
\usepackage{nomencl}    
\usepackage{xcolor,colortbl}
\usepackage{pbox}

\journal{Flow: Application of Fluid Mechanics}

\begin{document}

\begin{frontmatter}


\title{Large-eddy simulations to define building-specific similarity relationships for natural ventilation flow rates}


\sethlcolor{yellow}
\date{March 2022}

\author[stanford]{Yunjae Hwang}
\author[stanford]{Catherine Gorl\'e}

\address[stanford]{Department of Civil and Environmental Engineering,\\Stanford University, Stanford, CA 94305, USA}

\begin{abstract}
Natural ventilation can play an important role towards preventing the spread of airborne diseases in indoor environments. However, quantifying natural ventilation flow rates is a challenging task due to significant variability in the boundary conditions that drive the flow. In the current study, we propose and validate an efficient strategy for using computational fluid dynamics (CFD) to assess natural ventilation flow rates under variable conditions, considering the test case of a single-room home in a dense urban slum. The method characterizes the dimensionless ventilation rate as a function of the dimensionless ventilation Richardson number and the wind direction. First, the high-fidelity large-eddy simulation predictions are validated against full-scale ventilation rate measurements. Next, simulations with identical Richardson numbers, but varying dimensional wind speeds and temperatures, are compared to verify the proposed similarity relationship. Last, the functional form of the similarity relationship is determined based on 32 LES. Validation of the surrogate model against full-scale measurements demonstrates that the proposed strategy can efficiently inform accurate building-specific similarity relationships for natural ventilation flow rates in complex urban environments.
\end{abstract}

\end{frontmatter}

\clearpage
\section{Introduction}
Natural ventilation can play an important role towards preventing the spread of airborne diseases in indoor environments. The global COVID-19 pandemic has put a spotlight on the importance of ventilation, but in many low-income communities the use of natural ventilation could have significant benefits beyond the current pandemic. For example, a study in Dhaka, Bangladesh identified an association between the ventilation status of slum homes and the occurrence of pneumonia in children under five, which is the leading cause of death in this age group~\cite{wang2016global}. Specifically, households where pneumonia occurred were 28\% less likely to be cross-ventilated~\cite{ram2014household}.
 
To quantify the relationship between ventilation and health outcomes, it is essential to have an accurate estimate of the ventilation rate in a given home. Obtaining these estimates can be challenging when considering natural ventilation, since the flow rates through ventilation openings depend on the complex urban geometry, as well as on the highly variable driving forces due to wind and buoyancy~\cite{wang2017assessment, etheridge2011natural}. As a result, theoretical or empirical envelope models are likely to have limited accuracy when applied to configurations other than those for which they were derived or calibrated~\cite{karava2004wind,seifert2006cross,karava2007wind,caciolo2011full,karava2011airflow,larsen2018calculation}. 
The use of computational fluid dynamics (CFD) could further inform ventilation rate estimates by providing a detailed solution of the natural ventilation flow patterns and flow rates in a specific configuration, but two challenges remain to be addressed. 

The first challenge is the validation of CFD predictions of the complex flow phenomena that occur during combined buoyancy- and wind-driven natural ventilation in an urban environment. To date, validation of CFD results for natural ventilation has primarily focused on wind-driven ventilation processes, considering both small-scale~\cite{hu2008cfd, ramponi2012cfd, tominaga2016wind, van2017accuracy} and full-scale~\cite{jiang2002effect, larsen2011characterization} experiments. Buoyancy-driven ventilation has received comparatively less attention, possibly because the flow is more challenging to model. Slight changes in the thermal boundary conditions can lead to significant changes in the internal air-flow pattern~\cite{srebric2008bcs}, and the weak coupling between the momentum and energy equations can produce numerical stability issues~\cite{ji2007numerical}. Furthermore, it is difficult to achieve flow similarity 
in reduced scale experiments when heat transfer is involved~\cite{chen2009ventilationperformance, wykes2020effect}. As a result, validation for buoyancy-driven ventilation has primarily considered full-scale experiments~\cite{xing2001study, jiang2003buoyancy, bangalee2013computational}. Combined wind- and buoyancy-driven ventilation, where the two driving forces may produce assisting or opposing pressure gradients, is similarly challenging to model at reduced scale~\cite{wang2017assessment}. A full-scale validation study by Casciolo et al. considered single-sided ventilation in an isolated building~\cite{caciolo2011full, caciolo2012numerical}, but validation for cross-ventilation, or for a building in an urban environment, remains to be pursued.

The second challenge is the need to quantify the effect of the highly variable driving forces on the natural ventilation flow~\cite{Linden1999}. Combined wind- and buoyancy-driven ventilation is a high-dimensional problem, with the flow strongly affected by indoor surface and outdoor air temperatures, as well as wind conditions. The effect of variability in these conditions remains relatively unexplored, and evaluating a design under all possible conditions using CFD would be prohibitively expensive~\cite{etheridge2015}. Hence, CFD-based performance evaluations of natural ventilation systems will require some form of dimension reduction in the uncertain parameter space.  


The objective of this work is to propose and validate an efficient strategy for using CFD to predict natural ventilation flow rates as a function of highly variable outdoor weather and indoor thermal boundary conditions. To achieve this objective, we address the two outstanding challenges identified above, considering a test case of a representative home in an urban slum environment with natural ventilation through a window and a skylight. Given the important contribution of turbulence to the overall natural ventilation flow rate, CFD simulations are performed using the large-eddy simulation (LES) technique. First, we validate the LES predictions of the ventilation flow rate against field experiments, considering two measurements obtained under different boundary conditions. Second, we explore mapping the high-dimensional parameter space defining the variable boundary conditions onto two parameters: the ventilation Richardson number $Ri_v$, which represents the ratio of the driving forces due to buoyancy and wind, and the wind direction $\theta_{wind}$. We determine whether the dimensionless ventilation rate exhibits similarity in terms of $Ri_v$ by comparing predictions under different wind speeds and temperatures that correspond to identical $Ri_v$ values. Then, we perform 32 LES simulations to characterize the non-dimensional ventilation rate as a function of $Ri_v$ and $\theta_{wind}$. Last, the predictions obtained by the resulting surrogate model are compared to the ventilation rate measurements performed in the home. 

In the remainder of this paper, Section 2 introduces the test case and the corresponding field measurements. Section 3 discusses the LES set-up and Section 4 presents the results of the LES validation exercise. Section 5 introduces the proposed similarity relationship, including its validation. Section 6 presents the conclusions and areas for future research.

\section{Test case and field measurements} \label{sec:test_case_and_measurements}
On-site field measurements were conducted in a representative slum house for 15 days in February 2019. 
This section introduces the test house, and summarizes the field measurement setup and ventilation rate measurement technique.

\subsection{Description of the test house}\label{subsec:test_case}
\begin{figure}[htp!]
    \begin{subfigure}{0.45\textwidth}
    \centering
    \includegraphics[width=\textwidth]{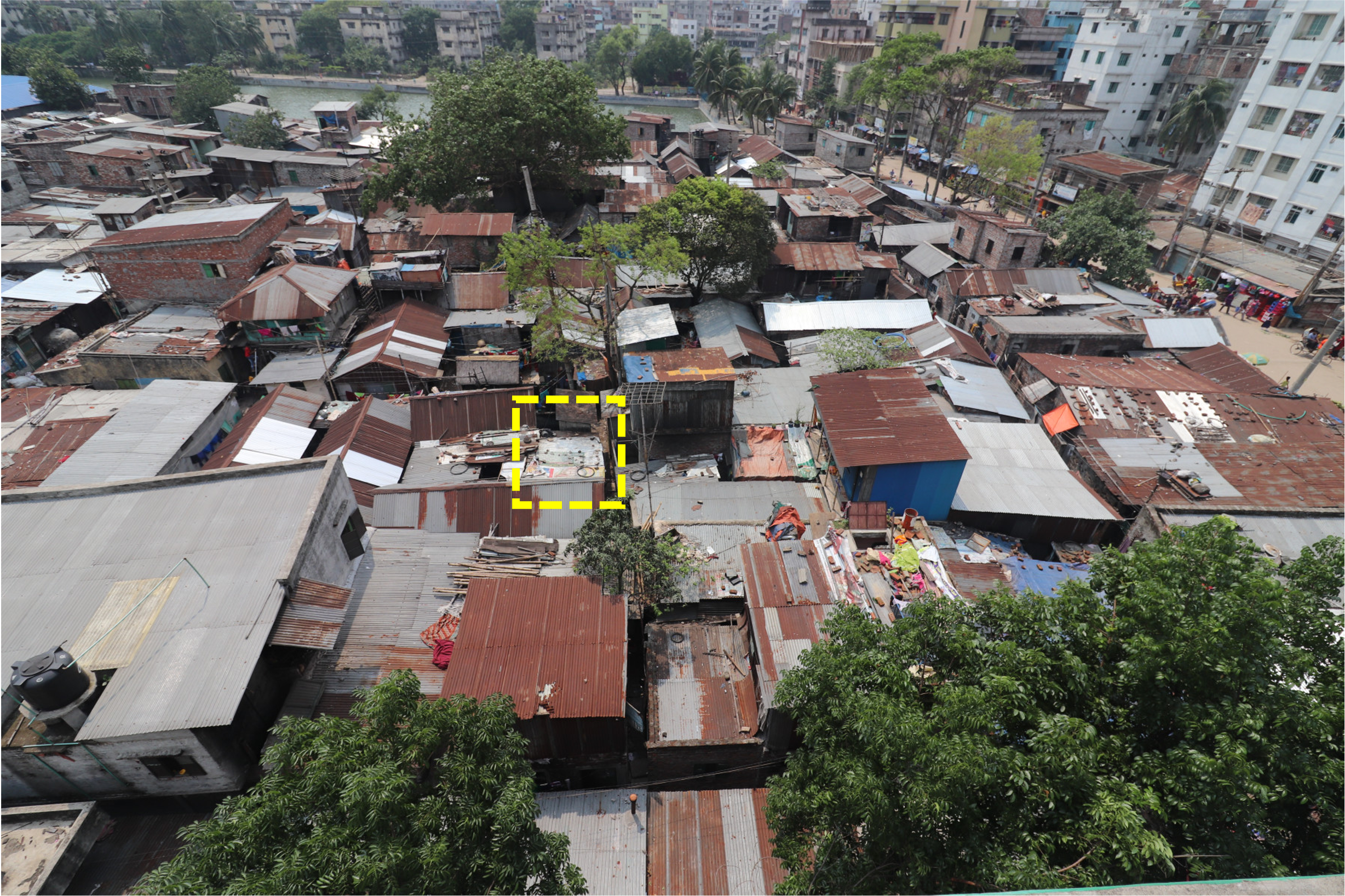}    
    \caption{ }
    \label{fig:outfall_slum}
    \end{subfigure}
    ~
    \centering
    \begin{subfigure}{0.45\textwidth}
    \centering
    \includegraphics[width=\textwidth]{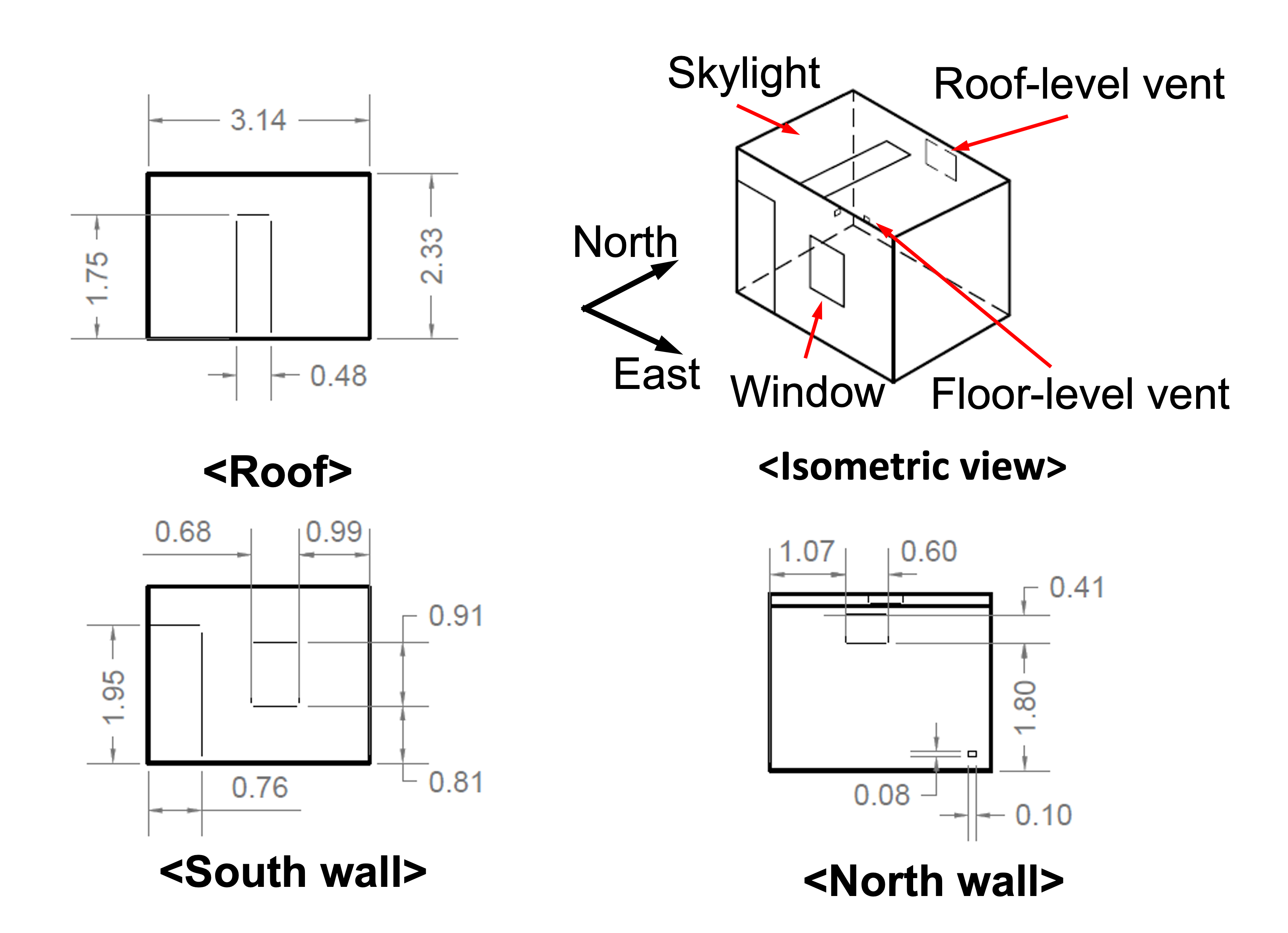}    
    \caption{ }
    \label{fig:outfall_house_drawing}
    \end{subfigure}
    \caption{Study site of the current research: (a) bird-eye view of area of interest, Bangladeshi urban-slum, and (b) drawing of the test house}
\end{figure}

The test house is a representative single-room home in Outfall, a low-income community in Dhaka, Bangladesh~\cite{islam2006slums}. The house has a rectangular floor plan and a slanted ceiling, shown in Figure~\ref{fig:outfall_slum}. The detailed dimensions are illustrated in Figure~\ref{fig:outfall_house_drawing}, where each wall is labeled based on its orientation. Multiple openings were constructed to determine the effectiveness of a variety of ventilation strategies: a skylight; a large window with a security grill on the south wall; a small floor-level vent on the north wall, and; a mid-size rear vent, also on the north wall. Four different configurations, each with two of the ventilation openings opened, were tested: (1) skylight and floor-level vent, (2) skylight and roof-level vent, (3) window and roof-level vent, and (4) skylight and window. The CFD modeling presented in this paper primarily focuses on the configuration with the skylight and the window. The skylight and floor-level vent configuration is considered once, to verify the proposed similarity relation in Section~\ref{sec:richardson_similarity}. 

\subsection{Temperature and wind measurements} \label{subsec:temp_and_wind_measurements}
The boundary conditions defining the natural ventilation flow were characterized based on temperature and wind measurements in the vicinity of the test house. 
Inside the house, 24 temperature sensors were installed: 15 thermistors measured the indoor air temperature at 
5 horizontal locations and 3 different heights, while 9 thermistors recorded the surface temperatures of the walls, the roof, and the floor. 
A mobile weather station was installed on the roof of the tallest building ($H_{max}$ = 25 m) in the area, collecting outdoor temperatures and free-stream wind velocities. Both temperature and wind data are recorded with a sampling frequency of 1Hz. 

\subsection{Ventilation rate measurements} \label{subsec:ach_measurements}
Ventilation rate measurements were performed in the test house with a tracer concentration decay technique. The experiments use particulate matter (PM) because of its low cost and widespread availability. Under ideal conditions, the ventilation rate can be determined from the exponential decay in the tracer concentration as time elapses:
\begin{equation} \label{eq:ach_equation}
    Q_{nv}(t) = V_{House} \cdot \frac{\log(c(t)) - \log(c_{peak})}{t - t_{peak}},
\end{equation}
where $t$ is the time, $c(t)$ is the concentration of the tracer at time $t$, $c_{peak}$ and $t_{peak}$ indicate the concentration and time of the peak, and $V_{House}$ is the volume of air in the house. The relationship assumes well-mixed conditions with a spatially uniform tracer concentration, as well as negligible values of the tracer in the outdoor environment. These conditions were challenging to achieve during the field experiments in the slum neighborhood, which introduces some uncertainty in the ventilation rates determined using the technique. To reduce and quantify this uncertainty, the signal is processed by first calculating a time series of $Q_{nv}$ using equation~\ref{eq:ach_equation}, and then computing the mean and the standard deviation of this time series, considering a 5- to 10-minute window with a quasi-steady state signal. By considering this quasi-steady state window, the effects of peaks observed at the start and end of some of the measured time series are eliminated. The standard deviation provides a measure of the fluctuations observed during this period of quasi-steady state decay. To facilitate a relative assessment of the ventilation status of the house independent of its specific volume, the ventilation rate will be presented in terms of air change per hour (ACH)
in the remainder of the manuscript, where $ACH(t) = Q_{nv}(t)/V_{House}$ in units 1/hr.

\section{Large-eddy simulations}\label{sec:LES}
The LES are performed using the CharLES solver~\cite{charles}. In this section, we introduce the governing equations, the setup of computational domain and mesh, and the inflow and other boundary conditions.
 
\subsection{Governing equations}
LES applies a filter to the instantaneous field quantities $u_i(x,t)$, splitting them into filtered $\widetilde{\langle\cdot\rangle}$ and sub-filter (sub-grid) components ${\langle\cdot\rangle}'$: $u_i(x,t)=\widetilde{u_i}(x,t)+u_i'(x,t)$. This results in the following equations for conservation of mass and momentum:
\begin{equation}\label{eq:continuity} 
  \frac{\partial \rho}{\partial t} + \frac{\partial \rho \widetilde{u}_j}{\partial x_j}=0
\end{equation}
\begin{equation}\label{eq:momentum} 
  \frac{\partial \rho \widetilde{u}_i}{\partial t}+ \frac{\partial \rho \widetilde{u}_i \widetilde{u}_j }{\partial x_j}  = -\frac{\partial \widetilde{p}}{\partial {x_i}} +\frac{\partial \widetilde{\sigma}_{ij}}{\partial x_j} + \rho g \delta_{i3}, 
\end{equation}
where $\widetilde{\sigma}_{ij}=(\mu+\mu_{sgs})(\frac{\partial \widetilde{u}_i}{\partial x_j} + \frac{\partial \widetilde{u}_j}{\partial x_i}-\frac{2}{3} \delta_{ij}\frac{\partial \widetilde{u}_k}{\partial x_k} )$ are the viscous and subgrid stresses. The subgrid stresses are modeled using the linear eddy viscosity assumption with the Vreman subgrid model to calculate $\mu_{sgs}$~\cite{vreman2004eddy}.
Considering the small temperature variations in our problem, the simulations apply the Boussinesq approximation to represent the effect of buoyancy, where density fluctuations due to temperature are neglected in the advection terms. The momentum source term due to buoyancy can then be replaced with: 
\begin{equation}\label{eq:boussinesq_approximation} 
    \rho g = \rho_{ref} g  \beta (\widetilde{T}-T_{ref}), 
\end{equation}
where $\rho$, $g$ and $\beta$ are the density, the gravitational constant and the thermal expansion coefficient, and the subscript \textit{ref} denotes the reference value of a quantity. The temperature field is obtained by solving the following filtered equation:
\begin{equation}\label{eq:heat} 
  \frac{\partial \rho\widetilde{T}}{\partial t}+  \frac{\partial\rho \widetilde{u}_j  \widetilde{T}}{\partial x_j} =  \frac{\partial}{\partial x_j}[(\frac{\widetilde{\alpha}}{c_p} + \frac{\mu_{sgs}}{Pr_{sgs}})\frac{\partial \widetilde{T}}{\partial x_j}],  
\end{equation}
where $\widetilde{\alpha}$, $c_p$, and $Pr_{sgs}$ are the thermal diffusivity, the specific heat capacity, and the subgrid Prandtl number, respectively. 

In addition to Eqs.~\ref{eq:continuity},~\ref{eq:momentum} and~\ref{eq:heat}, a filtered scalar transport equation is solved to mimic a tracer decay measurement and visualize the indoor ventilation pattern:
\begin{equation}\label{eq:scalar} 
    \frac{\partial\rho\widetilde{C}}{\partial t}
    +\frac{\partial\rho\widetilde{u}_j\widetilde{C}}{\partial x_j}
    =\frac{\partial}{\partial x_j} 
    \left[\left(\rho\widetilde{D}+\frac{\mu_{sgs}}{Sc_{sgs}}\right)\frac{\partial \widetilde{C}}{\partial x_j}\right],
\end{equation}
where $\widetilde{D}$ and $Sc_{sgs}$ are the viscous diffusion coefficient and the subgrid Schmidt number ($Sc_{sgs}=1$), respectively. 

\subsection{Computational domain and mesh}
\begin{figure}[htp!]    
\centering
    \begin{subfigure}{0.53\textwidth}
    \centering
    \includegraphics[width=\textwidth]{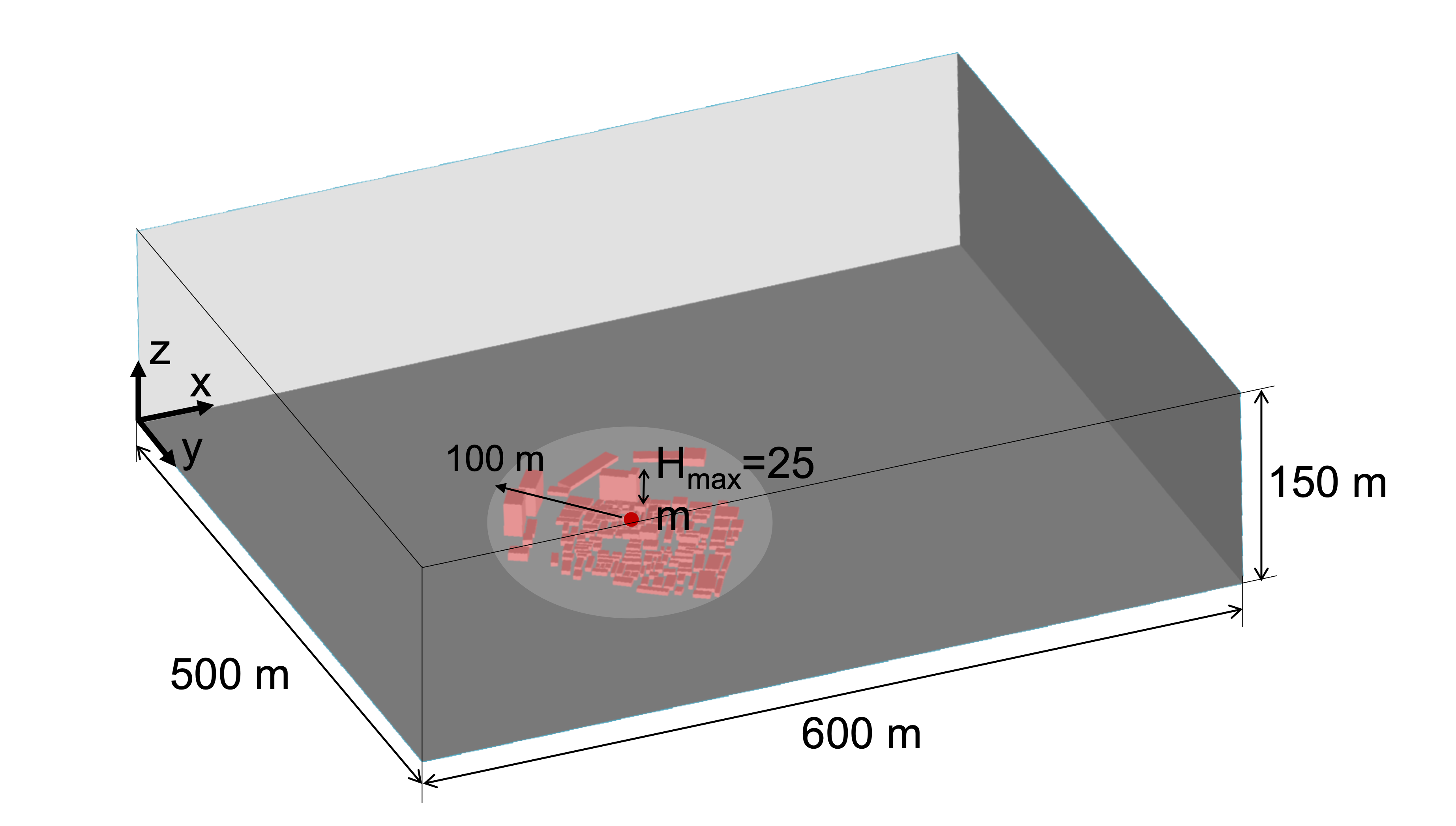}
    \caption{ }
    \label{fig:computational_domain}
    \end{subfigure}
    ~
    \begin{subfigure}{0.43\textwidth}
    \centering
    \includegraphics[width=\textwidth]{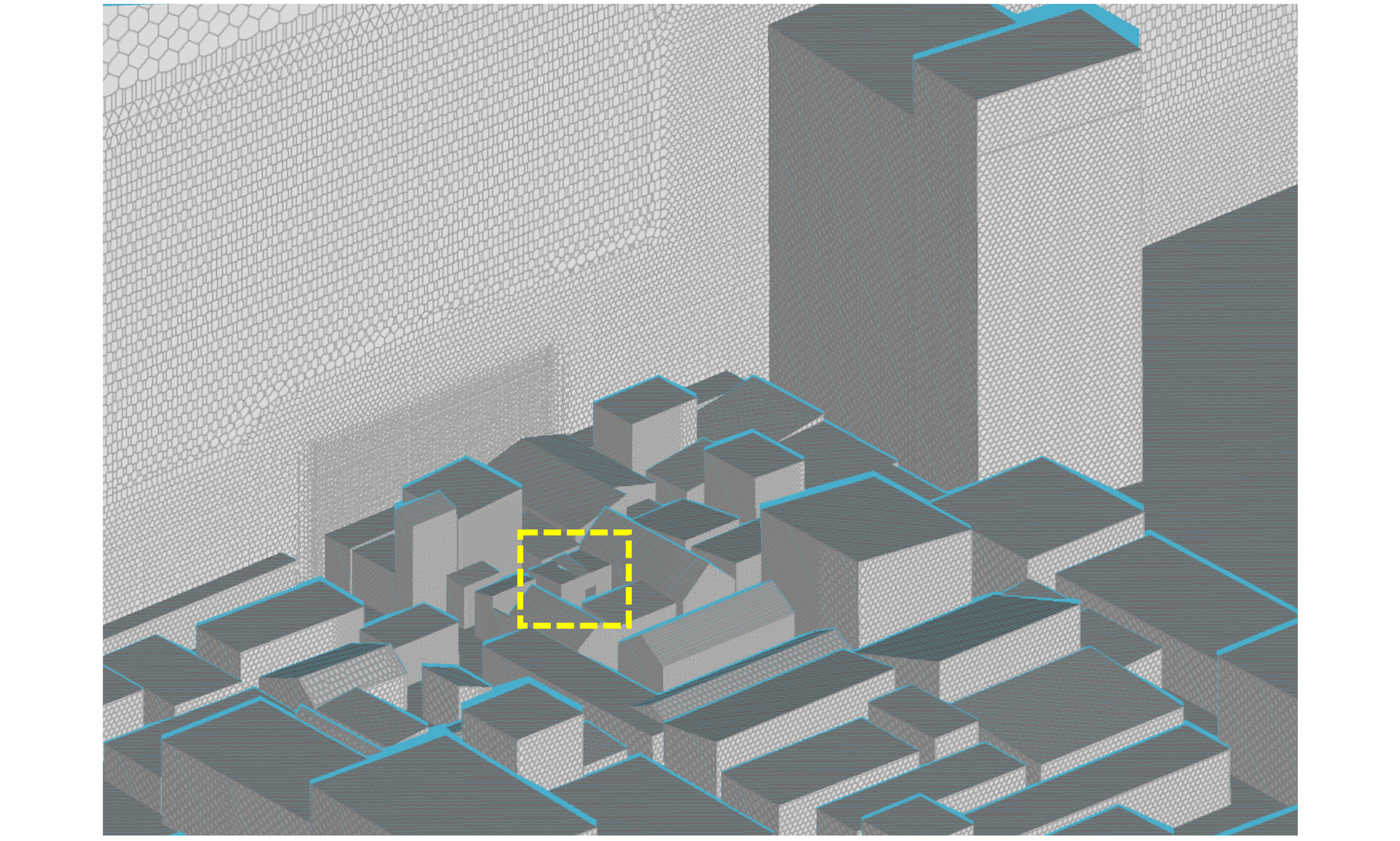}
    \caption{ }
    \label{fig:computational_mesh}
    \end{subfigure}
    \caption{Computational representation of area of interest: (a) computational domain and (b) mesh view in the building area}
    \label{fig:computational_domain_and_mesh}
\end{figure}

 Figure \ref{fig:computational_domain} shows the computational domain, centered around the test house. Our primary region of interest is the inside and the vicinity of the test house and the model includes an accurate representation of the buildings within a radius of $5H_{house}$. Buildings outside this immediate range but within 100 m are represented as rectangular blocks. The size of the domain was determined following best practice guidelines~\cite{franke2011cost}. The horizontal dimensions are $20H_{max}$ by $30H_{max}$, where $H_{max}$=25 m is the height of the tallest building in the domain. The inflow boundary is located at a distance greater than $5H_{max}$ from the most upstream building, while the outflow boundary is located $14H_{max}$ downstream of the test house. The vertical domain height is $6H_{max}$ and the lateral boundaries are at least $5H_{max}$ away from all building geometries. 

To consider different wind directions, the urban geometry is oriented differently inside the computational domain and a new grid is generated with the CharLES mesh generator. Each grid consists of approximately 21 million cells. A snapshot of a grid is shown in Figure~\ref{fig:computational_mesh}. The cell size ranges from 10 m in the background to 6.7 cm near the house. The refinement is introduced gradually using different refinement zones, and the resulting resolution adheres to the guidelines that are recommended for CFD models of natural ventilation and wind engineering applications \cite{tominaga2008aij, franke2011cost}. 
A grid sensitivity study showed that two finer computational grids predicted mean ventilation rates with negligible differences from the mesh used to generate the results presented in this paper.

\subsection{Boundary conditions}\label{subsec:LES_bcs}
For the turbulent inflow condition, we use the divergence-free version of a digital filter method developed for wind engineering applications~\cite{xie2008efficient, kim2013divergence}. The method generates an unsteady inflow with turbulence structures that are coherent in space and time, based on input for the mean velocity and Reynolds stress profiles, and for the turbulence length scales. A limitation of the digital filter inflow generation method is that the turbulence tends to decay as the flow moves through the domain, such that the turbulence intensities at the location of interest may be considerably lower than those specified at the inlet. To resolve this issue we employ a gradient-based optimization technique, where the Reynolds stress profiles and length scales at the inflow boundary are optimized to obtain the desired target profiles just upstream of the first row of buildings in the domain~\cite{lamberti2018optimizing}.

\begin{figure}
    \centering
    \includegraphics[width=0.8\textwidth]{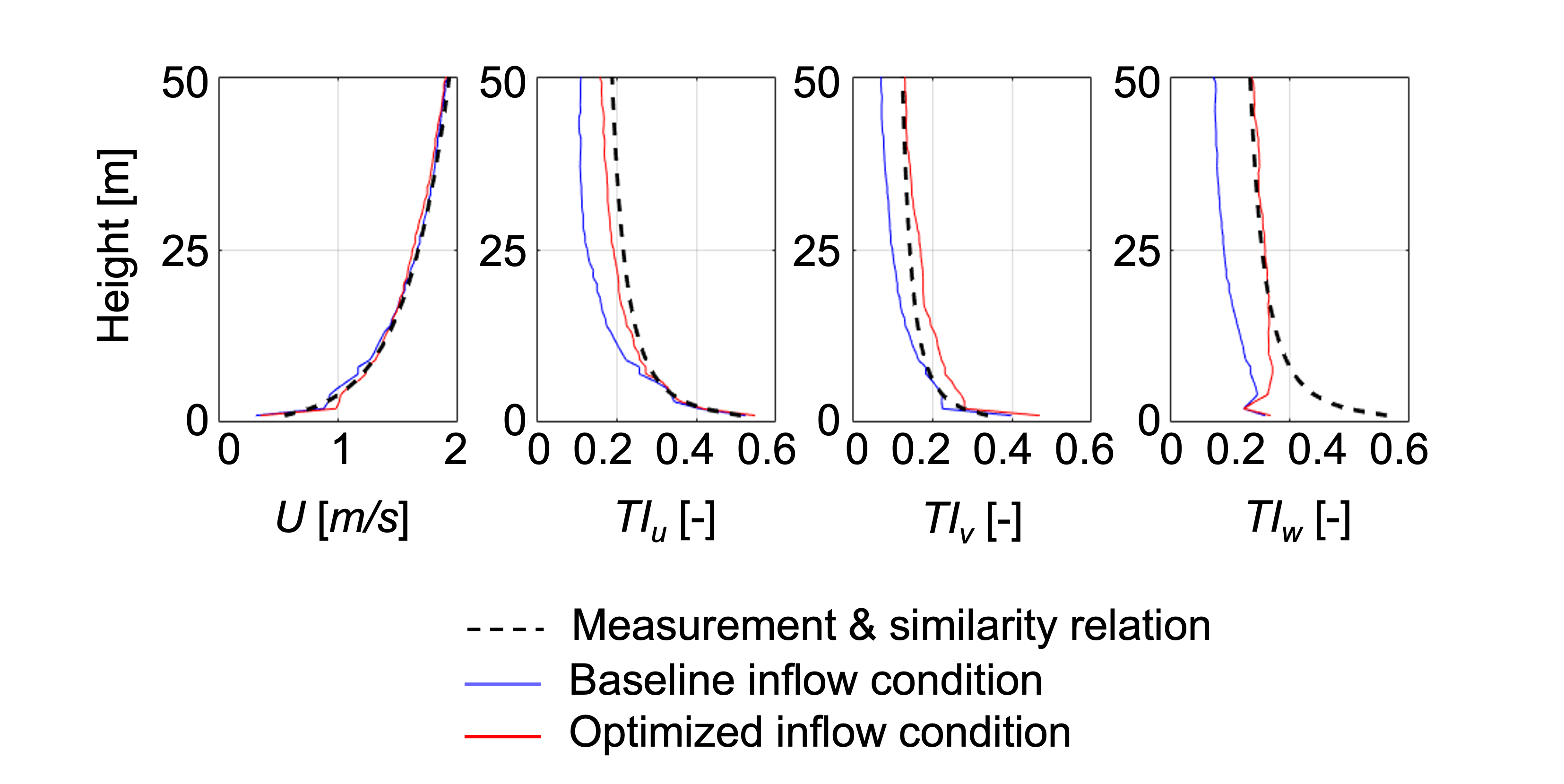}
    \caption{Inflow and target profiles for the LES validation study}
    \label{fig:inflow_optimization}
\end{figure}

Figure~\ref{fig:inflow_optimization} presents the target profiles and the corresponding optimized inflow profiles used for the validation study (Section~\ref{sec:LES_validation}). 
The mean velocity corresponds to a logarithmic profile:
\begin{equation} \label{eq:log_law}
     U(z) = \frac{u_*}{\kappa} \log ( \frac{z+z_0}{z_0}), 
 \end{equation}
 where $u_*$ is the friction velocity, $\kappa$ is the von-Karman constant (0.41), and $z_0$ is the roughness length. The roughness length is set to 0.5 m, corresponding to urban terrain~\cite{wieringa1992updating}). The friction velocity is calculated from the freestream velocity $U_{wind}$ at $25$ m height. For the Reynolds stresses, the profiles are obtained from similarity relationships~\cite{stull2012abl}:
 \begin{equation} \label{eq:InletTI}
     \overline{u'u'} = 5.7 u_*^2; \quad 
     \overline{v'v'} = 2.8 u_*^2; \quad
     \overline{w'w'} = -\overline{u'v'} = 2.5 u_*^2.
 \end{equation}
Lastly, the turbulence length scales are estimated using the free stream velocity measurements. The auto-correlation of the stream-wise velocity component indicates a time scale of $\tau_u$=15 s, which is converted to a length scale of $^xL_u$=25 m using Taylor's hypothesis. The remaining 8 length scales are estimated as a fraction of $^xL_u$~\cite{emes2018estimating}: 
\begin{align*} \label{eq:length_scales}
&^xL_u=1.00^xL_u,\quad ^xL_v=0.20^xL_u,\quad ^xL_w=0.30^xL_u; \\ 
&^yL_u=0.28^xL_u,\quad ^yL_v=0.32^xL_u,\quad ^yL_w=0.07^xL_u; \\
&^zL_u=0.27^xL_u,\quad ^zL_v=0.14^xL_u,\quad ^zL_w=0.06^xL_u.\\
\end{align*}

The outlet boundary condition is a zero gradient condition. 
At the ground and building surfaces wall functions are applied. A rough-wall function for a neutral atmospheric boundary layer with $z_0$=0.5 m is imposed at the ground boundary, while a standard smooth wall model is used for the building surfaces. The two lateral boundaries are periodic and a slip condition is applied at the top boundary. 

For the thermal boundary conditions, we impose a constant temperature for the indoor wall, roof, and floor surfaces in the test house. Adiabatic conditions are used for the outdoor ground and the surrounding building walls. A constant temperature is also specified at the inflow boundary such that a quasi-steady state solution with a constant temperature difference between indoor and outdoor will be reached. 

The simulations presented in this paper consider a variety of weather conditions in terms of the free stream wind speed, wind direction, and the indoor surface and inflow temperatures. For the validation presented in Section~\ref{sec:LES_validation}, the values were specified based on the specific field measurements being modeled. For the similarity analysis in Section~\ref{sec:richardson_similarity}, the range of likely conditions during the entire winter season is considered.

\subsection{Discretization methods and solution procedure}
The computational domain is discretized using hexagonal close-packed cells created by the solver's built-in mesh generating tool. 
The solver uses a second-order central discretization in space as well as second-order implicit time advancement with a fixed time-step size. Given a fixed resolution of the computational grid, different combinations of wind and temperature boundary conditions result in different indoor ventilation and outdoor wind flow patterns. Depending on the specific conditions, the time-step is chosen in the range of 0.001 s to 0.05 s such that the resulting maximum CFL number is less than 1.0. Statistics of the quantities of interest are calculated from flow solutions obtained over 150 $\tau_{ref}$, after an initial burn-in period of at least 100 $\tau_{ref}$, where $\tau_{ref}=L_{House}/U_{wind}$ is the flow-through time over the test house, i.e. the ratio of the length scale of the house to the wind speed at the reference height.


\subsection{Calculation of the ventilation rate from LES} \label{subsec:calculation_vent_rate}
Ventilation rates will be calculated from the LES using two approaches.
The first approach uses a tracer concentration decay techniques, similar to the field experiment. The additional equation for scalar transport is solved, with the scalar field initialized to a constant non-zero value inside the test house and a zero value outside. The scalar concentration decay at the center of the house is recorded to compute the ventilation rate using equation~\ref{eq:ach_equation}. The main difference with the field measurement is that the simulated flow field does not correspond to a still environment at the start of the scalar concentration decay; the scalar is initialized once the burn-in period for the simulation of the velocity and temperature fields has passed and a quasi-steady state condition is reached. This approach provides an estimate of the air exchange rate at the monitored locations. However, this estimate is not necessarily an accurate representation of the indoor/outdoor air exchange rate, especially when the space is not uniformly ventilated. 

The second approach provides a more direct measure of this overall ventilation rate by calculating the instantaneous net amount of air exchange as half of total airflow through the two openings:
\begin{equation} \label{eq:vent_rate_velocity}
    Q_{nv, velocity} (t) = \frac{1}{2} ( \int |\mathbf{u}(t) \cdot \mathbf{n}_{1}| \,d\text{A}_1 + \int |\mathbf{u}(t) \cdot \mathbf{n}_{2}|\,d\text{A}_2 ),
\end{equation}
where $\mathbf{u}(t)$ is the instantaneous velocity field, and $\mathbf{n}$ and $d\text{A}$ are the normal vector and area of the opening denoted with the subscript 1 and 2. This calculation is similar to the cumulative average instantaneous ventilation rate introduced by Jiang et al. for single-sided ventilation~\cite{jiang2001study}. 

For the validation study in Section~\ref{sec:LES_validation}, the ACH values obtained with both approaches will be presented. When developing the similarity relationship in Section~\ref{sec:richardson_similarity}, we will use the ACH calculated by integrating the velocity at the openings, which is the most commonly adopted approach in ventilation studies using CFD.



\section{Validation of LES for predicting ACH }\label{sec:LES_validation}
Validation of the ACH predictions is performed for two different ventilation measurements. The following section first summarizes the measurement conditions and the corresponding boundary conditions for the LES. Subsequently, the LES results and the comparisons to the measured ACH values are presented.

\subsection{Ventilation measurements used for validation}
The two ventilation rate measurements used for validation were performed for the ventilation configuration with the skylight and window open. The measurements include one daytime and one nighttime experiment, to represent different conditions in terms of the combination of the driving forces due to wind and buoyancy. The resulting operating conditions are summarized in Table~\ref{tab:validation_measurements}. The wind speed and direction are almost identical to each other; hence, the difference in the ventilation rates can be attributed to the different temperature conditions. During the day, the outdoor air temperature is higher than the volume-averaged indoor air temperature and the indoor environment is thermally stratified. During the night, the outdoor temperature is lower than the indoor temperature, and the indoor temperature is relatively uniform. The measured wind conditions, as well as the outdoor air temperatures, $T_{outdoor}$, and the surface temperatures, $T_{roof}$, $T_{wall}$ and $T_{floor}$, are used to define the boundary conditions in the simulations as introduced in 
Section~\ref{subsec:LES_bcs}. 

\begin{table}[ht!]
    \centering
    \begin{tabular}{|l|r|r|} \hline
                    &  \textbf{Daytime} & \textbf{Nighttime} \\
                    & Feb. 7. 2019      & Feb. 11. 2019     \\ \hline \hline
Wind speed [m/s]            &   1.69    & 1.71              \\ \hline 
Wind direction [$^\circ$]   &   334     & 333               \\ \hline 
$T_{outdoor}$ [$^\circ$C]   &  \cellcolor{red!10} 28.35    & \cellcolor{blue!10} 16.15             \\ \hline 
$T_{indoor}$ (volume-averaged) [$^\circ$C]    &  \cellcolor{red!10} 26.50    & \cellcolor{blue!10} 20.15             \\ \hline 
$\quad T_{roof}$ [$^\circ$C]  & \cellcolor{red!10} 30.35 & \cellcolor{blue!10} 15.95   \\ \hline 
$\quad T_{wall}$ [$^\circ$C]  & \cellcolor{red!10} 25.00 & \cellcolor{blue!10} 20.55   \\ \hline 
$\quad T_{floor}$[$^\circ$C]  & \cellcolor{red!10} 21.60 & \cellcolor{blue!10} 21.15   \\ \hline 
\bf{ACH, measurements} [1/hour]   & \bf{9.65$\pm$0.66} & \bf{16.17$\pm$1.49}   \\ \hline 
    \end{tabular}
    \caption{Wind and temperature boundary conditions of two validation cases and their measurement results}
    \label{tab:validation_measurements}
\end{table}

\subsection{LES results for the velocity, temperature, and scalar fields} 
\begin{figure}
    \centering
    \includegraphics[width=0.8\textwidth]{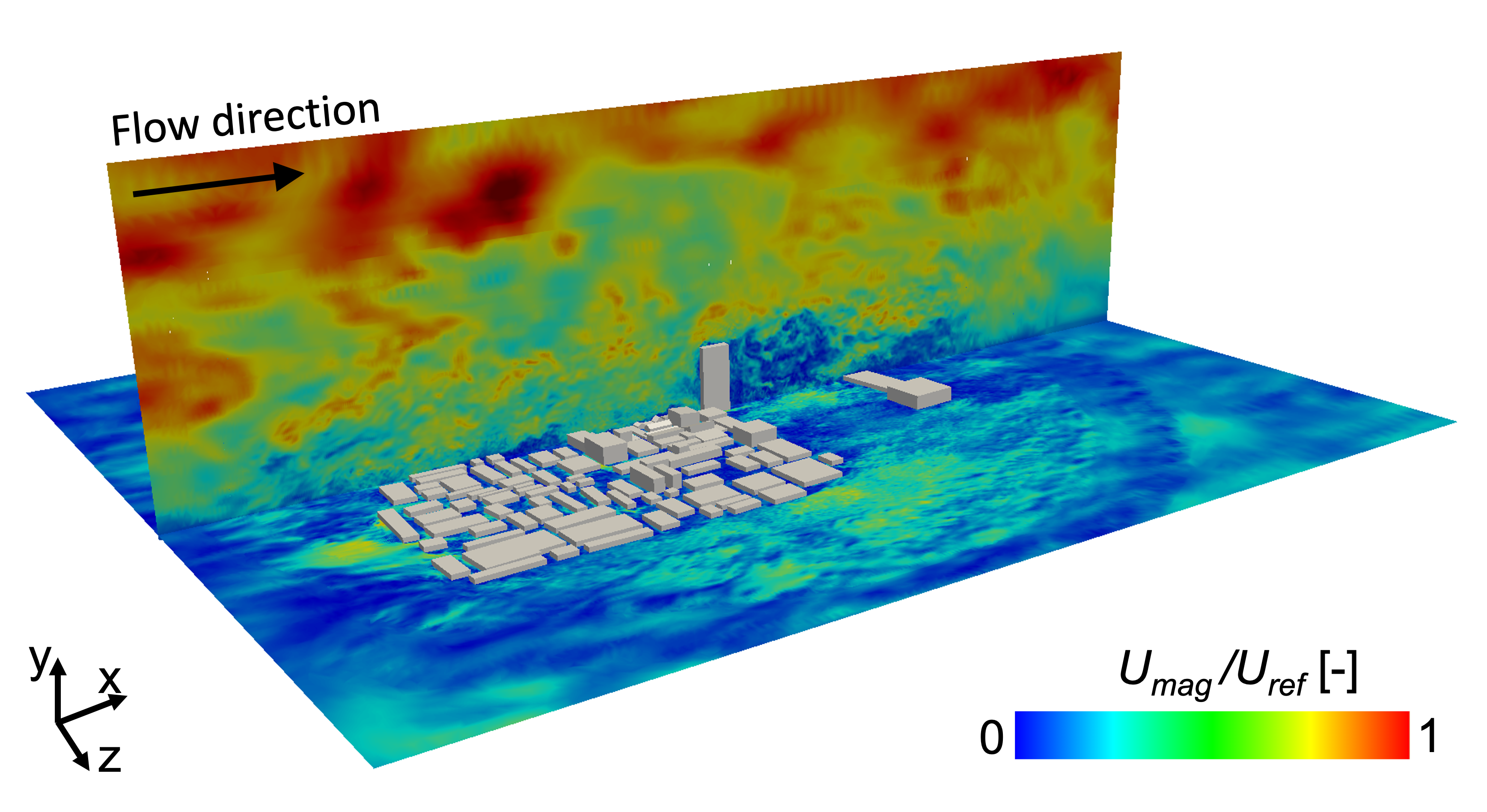}
    \caption{Instantaneous contours of the velocity magnitude for a vertical plane at the center of the domain and a horizontal plane at 1m from the ground }
    \label{fig:flow_field}
\end{figure}

Figure~\ref{fig:flow_field} visualizes the overall flow field using contour plots of the instantaneous velocity magnitude on a vertical plane through the center of the domain and on a horizontal plane at 1m above ground level. The contour plots visualize the large-scale turbulence structures in the boundary layer as well as the complexity of the flow within the urban canopy. 

\begin{figure}[htp!]
    \centering
    \includegraphics[width=1.0\textwidth]{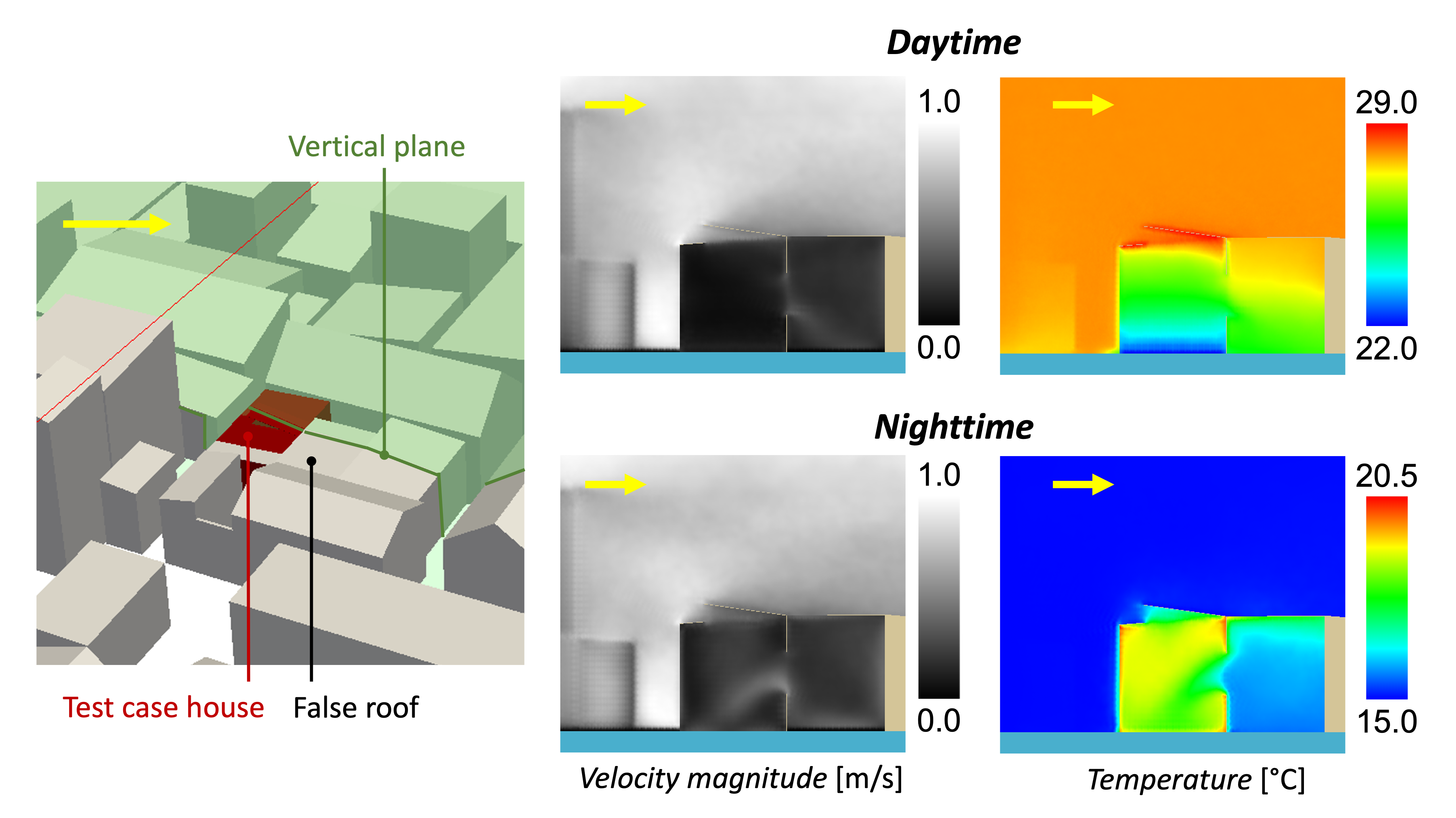}
    \caption{Contour plots of the time-averaged velocity magnitude (left) and temperature (right) on a vertical plane through the center of the house}
    \label{fig:validation_vel_temp}
\end{figure}

Figure~\ref{fig:validation_vel_temp} focuses on the flow in the test house, displaying contours of the time-averaged velocity and temperature fields on a vertical plane through the center of the house for the daytime (top) and nighttime (bottom). The daytime case has a low mean indoor air velocity, which indicates that ventilation will mainly be due to turbulent air exchange. The nighttime case exhibits a more pronounced mean flow through the window, indicating a potentially higher ventilation rate. The differences between the daytime and nighttime velocity patterns are related to the differences in the temperature fields. During the day the roof is heated by solar radiation, thereby establishing a strong vertical temperature stratification. During the night the temperature is much more uniform throughout the indoor space.

\begin{figure}[htbp]
    \centering
    \includegraphics[width=\textwidth]{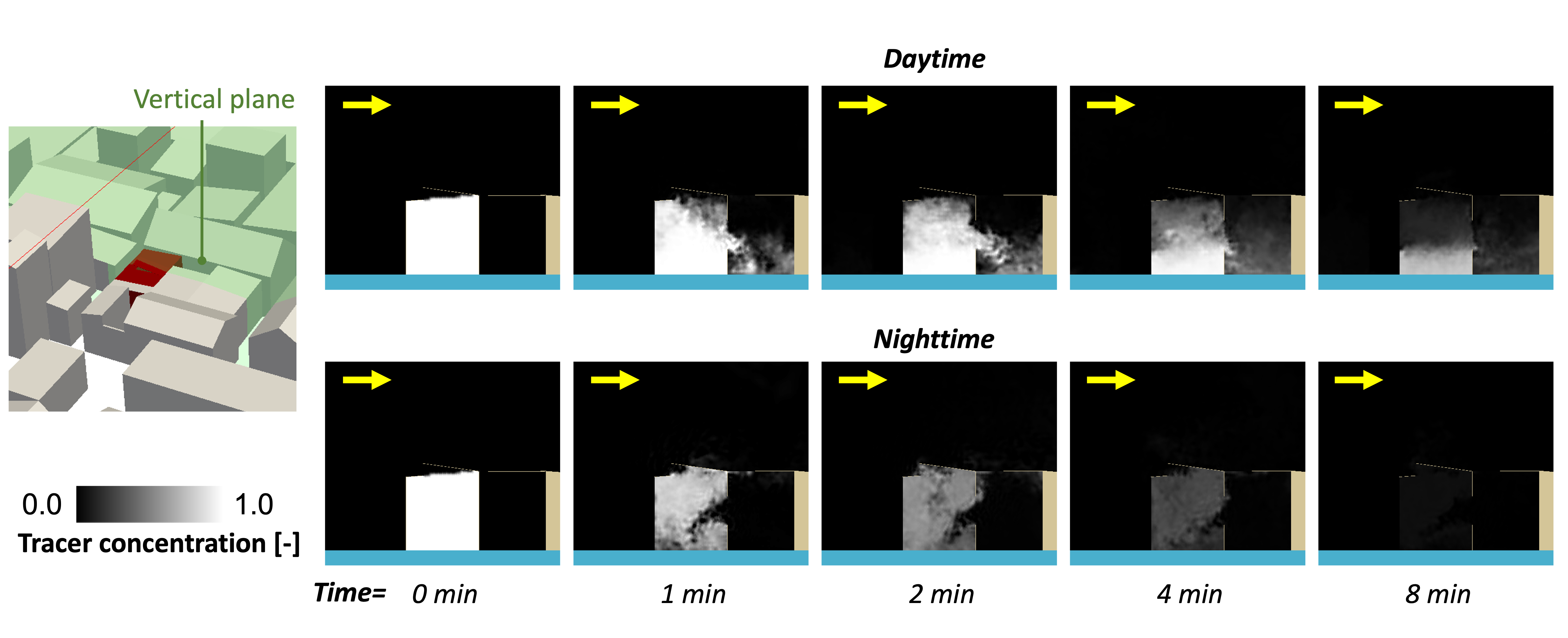}
    \caption{Time evolution of the scalar field after uniform initialization inside the house}
    \label{fig:validation_scalar}
\end{figure}

Figure~\ref{fig:validation_scalar} visualizes how these differences in the temperature distribution result in very different ventilation patterns by showing the time evolution of the scalar field on a vertical plane crossing both the window and skylight openings. During the daytime (top row), the stable temperature stratification results in a highly non-uniform ventilation of the indoor space, with the scalar concentration below the window opening much higher than the concentration near the ceiling. At night (bottom row), the more uniform temperature results in a more uniformly ventilated indoor space. 

\subsection{Validation of ACH predictions} 

\begin{figure}[htb]  
\centering
\begin{subfigure}{\textwidth}
    \centering
    \includegraphics[width=\textwidth]{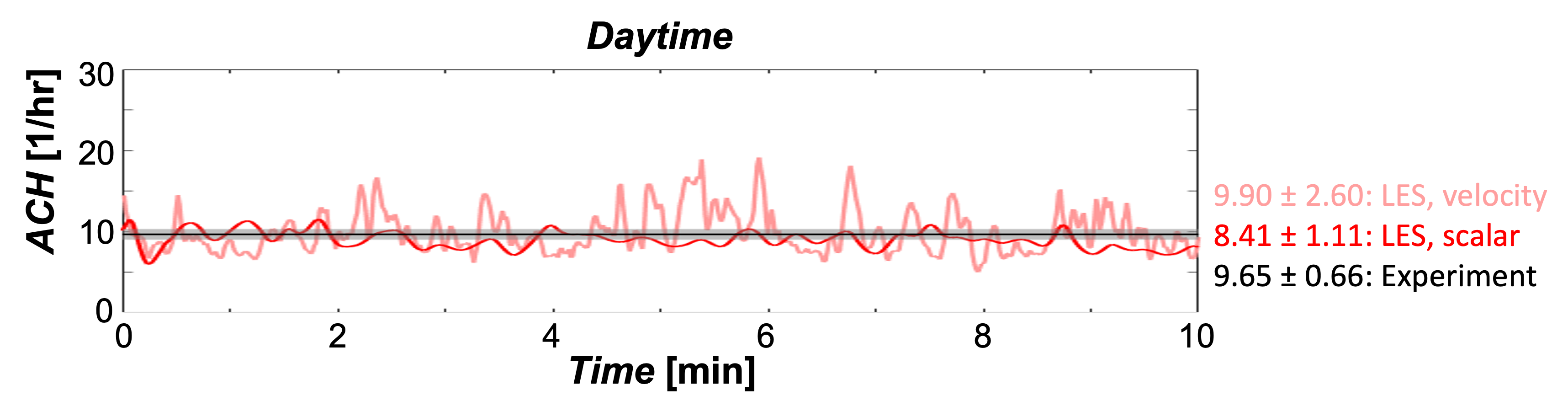}
\end{subfigure}
~
\begin{subfigure}{\textwidth}
    \centering
    \includegraphics[width=\textwidth]{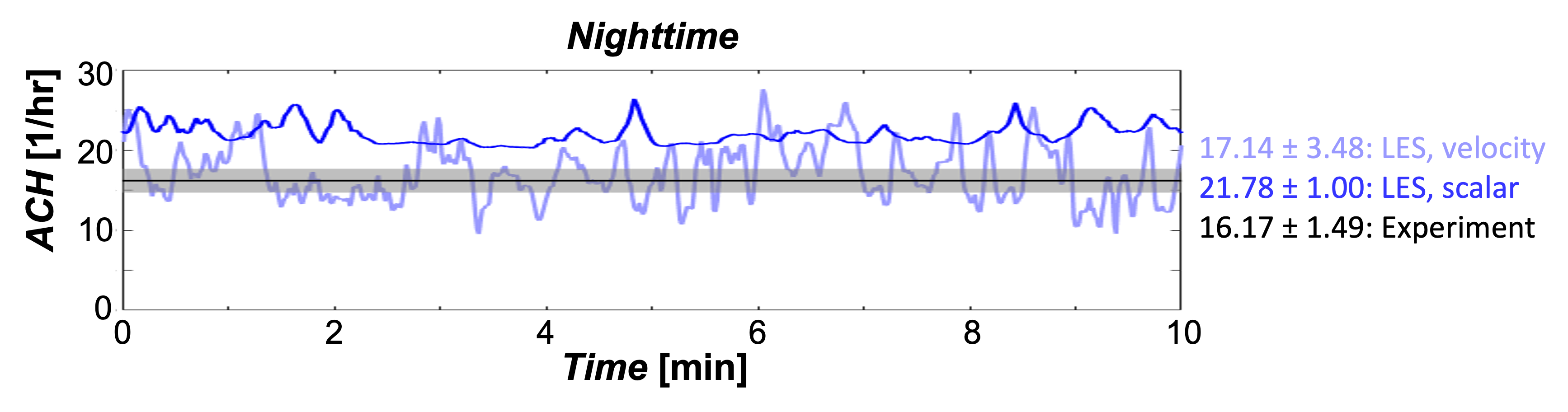}
\end{subfigure}
~
\caption{Time series of ACH estimated using the scalar concentration decay method and the velocity integration method for daytime (a) and nighttime (b) validation cases. Comparison to the mean and standard deviation of the ACH value obtained from the field measurement.}
\label{fig:ACH_validation_scalar}
\end{figure}
As introduced in Section~\ref{subsec:calculation_vent_rate}, ventilation rates are calculated using both the scalar concentration decay method, and the velocity integration method. Figure~\ref{fig:ACH_validation_scalar} presents the resulting comparison between the field measurements and the LES predictions for both the daytime and nighttime cases. 
Considering the daytime case, the values obtained from the field measurement and the LES agree very well. The difference between both LES estimates is ~16\%, with values of 8.41 1/hr and 9.90 1/hr for the scalar decay and the velocity integration methods respectively. The experimental value is in between both estimates at 9.65 1/hr. Despite the non-uniformity of the scalar field under daytime conditions, the scalar decay method provides a good estimate of the overall air exchange rate. This good agreement can be attributed to the central location at which the scalar decay was monitored; locations closer to the ceiling would over predict the ventilation rate, while closer to the ground an under prediction would be obtained. The velocity integration method reveals significant fluctuations over time, confirming the importance of unsteady, turbulent air exchange through the openings.

Considering the nighttime case, the difference between both predictions is slightly higher at ~24\%, with values of 21.78 1/hr and 17.14 1/hr for the scalar decay and the velocity integration methods respectively. The measurement produced a slightly lower value at 16.17 1/hr. There are two likely explanations for these observed differences. First, even though the space is more uniformly ventilated during the night than during the day, the result of the concentration decay method can still be very sensitive to the location at which the scalar decay is monitored. Small differences in the internal flow pattern, such as a slight shift in the direction of the outdoor air stream coming in through the window can result in non negligible variations in the decay rate (see Fig.~\ref{fig:validation_scalar}). 
Second, the simulations do not account for infiltration through small gaps in the building envelope. Infiltration can lead to additional air exchange, and it can also decrease the pressure differences between the indoor and outdoor environment, which can further modify ventilation pattern.

The above comparison of the ventilation rates demonstrates the predictive capability of LES for combined wind- and buoyancy-driven ventilation in a complex urban environment. The LES predicts measured ventilation rates within 24\%, and the simulations reproduce the significant difference between daytime and nighttime ventilation rates. In the next section, we leverage the validated LES setup to perform predictive simulations under different weather conditions and investigate whether flow similarity can be leveraged to efficiently characterize the ventilation in the home under a wide range of weather conditions. 

\section{Richardson number similarity for natural ventilation} \label{sec:richardson_similarity}
This section first introduces ventilation Richardson number similarity for natural ventilation. Subsequently, results from two simulations that have the same ventilation Richardson number but different operating conditions in terms of the wind speed and indoor/outdoor temperature difference are compared to confirm their similarity. Finally, simulations for a range of ventilation Richardson number and wind directions are performed to establish a similarity relationship that can efficiently account for the variability in weather conditions.

\subsection{Ventilation Richardson number, $Ri_v$} \label{subsec:ventilation_richardson_number}
Natural ventilation is driven by two driving forces, i.e., buoyancy and wind. Hence, the natural ventilation rate in a home will be a function of a large number of parameters affecting these two driving forces. For a fixed urban setting and natural ventilation opening configuration, one can expected the following dependency: 
\begin{equation} \label{eq:vent_rate_general_form} 
    Q_{nv} = f(T_{in}, T_{out}, T_{roof}, T_{wall}, T_{floor}, U_{wind}, \theta_{wind}, H, g, A_{opening}),
\end{equation}
where $T_{in}$ and $T_{out}$ are the indoor and outdoor air temperatures, $T_{roof}$, $T_{wall}$ and $T_{floor}$ are the roof, wall, and floor surface temperatures, $U_{wind}$ and $\theta_{wind}$ are wind speed and direction, $H$ is the height of the house, $g$ is the gravitational acceleration, and $A_{opening}$ is the effective area of the natural ventilation openings. The inherent variability in these parameters, due to varying weather conditions and indoor heat gains, makes it challenging to design natural ventilation systems~\cite{boulard2002bcs, jiang2002effect, srebric2008bcs, van2010effect, liu2019cfd}. 

\begin{figure}[htb]
    \centering
    \includegraphics[width=\textwidth]{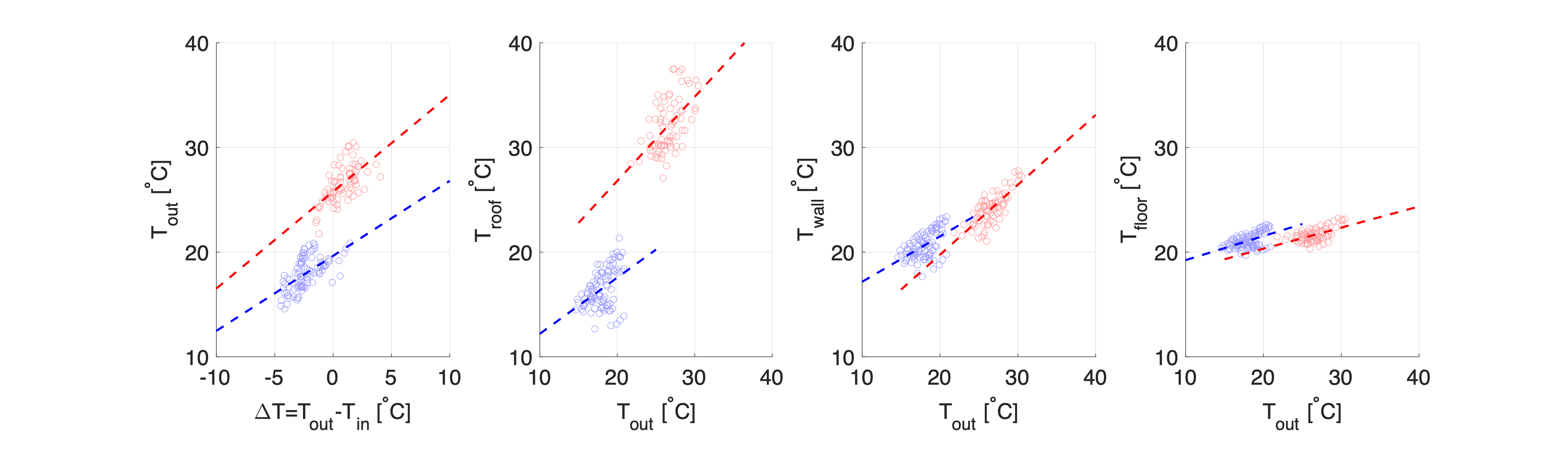}
    \caption{Correlation between measured temperatures (circles) and best linear fit (dashed lines) during daytime (red) and nighttime (blue)}
    \label{fig:temperature_correlation}
\end{figure}

As a first step towards reducing the dimensionality of the problem, we consider that for a certain construction of a home, the different temperatures are likely to be correlated. This is demonstrated in Figure~\ref{fig:temperature_correlation}, which shows scatter plots of hourly temperature measurements during daytime (red) and nighttime (blue) together with the best linear fit. 
The plots indicate that $T_{out}$ is correlated with $\Delta T = T_{out}-T_{in}$, $T_{roof}$, $T_{wall}$ and $T_{floor}$. Hence, for a given outdoor temperature, wind speed, and wind direction, the ventilation rate in a specific house can be calculated using a single simulation. This dependency is equivalent to the dependency that generally appears in analytical or empirical envelope flow models~\cite{warren1978ventilation,de1982ventilation, warren1984singlesided,hunt1999fluid,larsen2008single, cen2017revised}:
\begin{equation} 
\label{eq:vent_rate_dimensional} 
    Q_{nv} = f(\Delta T / T_{ref} \cdot g \cdot H , U_{wind}, \theta_{wind},A_{opening}).
\end{equation}

To further reduce the dimensionality of the problem, we define the dimensionless natural ventilation flow rate, as well as the dimensionless ventilation Richardson number $Ri_v$, which quantifies the ratio of the driving forces due to buoyancy and wind:
\begin{equation} \label{eq:Richardson_number} 
   Q'_{nv} = \frac{Q_{nv}}{A_{opening}\cdot U_{wind}} \quad;\quad Ri_v = \frac{\Delta T / T_{ref} \cdot g \cdot H }{ U_{wind}^2}.
\end{equation} 
The non-dimensional ventilation rate can then be written as a function of only two input parameters: 
\begin{equation} \label{eq:nondim_vent_rate}
    Q'_{nv}= \phi(Ri_v, \theta_{wind}).
\end{equation}
The main assumption in this similarity relationship is that, for a constant $Ri_v$, small changes in the non-dimensional indoor temperature field, caused by the different correlations of the floor, wall, and roof surface temperatures with $\Delta T/T_{out}$, will have a limited effect on $Q'_{nv}$. This assumption will be verified in the following section, before identifying the functional form of $\phi$ in Section~\ref{subsec:similarity_predictions}.

\subsection{Verification of the use of similarity relation} \label{subsec:similarity_verification}
The $Ri_v$ similarity proposed in Section~\ref{subsec:ventilation_richardson_number} is verified by performing simulations for two different ventilation configurations: daytime ventilation in the skylight/window configuration, and nighttime ventilation in the skylight/floor-level vent configuration. For each of these ventilation scenarios, two simulations with the same $Ri_v$ and $\theta_{wind}$, but different indoor-outdoor temperature differences ($\Delta T$) and reference wind speeds ($U_{wind}$) are performed. For the reference case, the wind and outdoor temperature boundary conditions are based on the field measurements; the measured $\Delta T$ and $T_{out}$ are used to determine the corresponding $Ri_v$. 
For the similar case, a different wind speed is selected, and the corresponding $\Delta T$ is calculated such that the two new parameters result in the same $Ri_v$. 
The outdoor temperature ($T_{out}$) and indoor surface temperature boundary conditions ($T_{roof}$, $T_{wall}$, and $T_{floor}$) are obtained using the correlations between the $\Delta T$ and these wall surface temperatures, shown in Figure~\ref{fig:temperature_correlation}. It is noted that $\Delta T$, and hence the actual $Ri_v$, are ultimately outputs of the simulations, i.e. their actual values depend on the indoor temperature calculated by the simulation. For each simulation it was verified that the difference between the intended and actual $Ri_v$ was negligible.

\begin{table}[ht!]
    \centering
    \begin{tabular}{|l|r r| r r|} \hline
  & \includegraphics[width=0.12\textwidth]{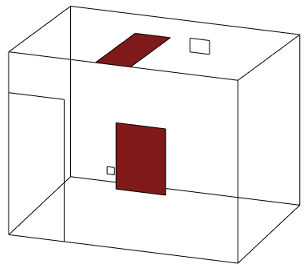} & \pbox{4cm}{Skylight \\ /window } & \includegraphics[width=0.12\textwidth]{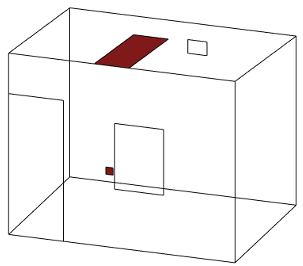} & \pbox{4cm}{Skylight \\ /floor-level vent }\\
  & \textbf{Daytime} & \textbf{Similar}     & \textbf{Nighttime} & \textbf{Similar} \\ \hline \hline
\textbf{Richardson \# [-]} & \textbf{0.052} & \textbf{0.052} &  \textbf{-0.041} & \textbf{-0.041}  \\ \hline 
Wind direction [$^\circ$]   &  330  &  330  &   330   &  330    \\ \hline 
Wind speed [m/s]            &  1.69 &  2.7  &   2.7   &  1.71   \\ \hline 
$\Delta T$ [$^\circ$C]      &  1.85 &  4.87 &  -3.90  & -1.52   \\ \hline \hline 
\textbf{mean($Q'_{nv}$) [-]} \cellcolor{gray!10} & \cellcolor{gray!10} \textbf{0.0200} & \cellcolor{gray!10} \textbf{0.0187} & \cellcolor{gray!10} \textbf{0.0169} & \cellcolor{gray!10} \textbf{0.0175} \\ \hline
\textbf{std($Q'_{nv}$)  [-]} \cellcolor{gray!10} & \cellcolor{gray!10} \textbf{0.0051} & \cellcolor{gray!10} \textbf{0.0060} & \cellcolor{gray!10} \textbf{0.0053} & \cellcolor{gray!10} \textbf{0.0054} \\ \hline
\end{tabular}
    \caption{Summary of operating conditions ($Ri_v$, $\theta_{wind}$, $U_{wind}$ and $\Delta T$) and simulation results for the verification of Richardson number similarity}
    \label{tab:Ri_similarity_verification}
\end{table}

\begin{figure}
    \centering
    \includegraphics[width=0.9\textwidth]{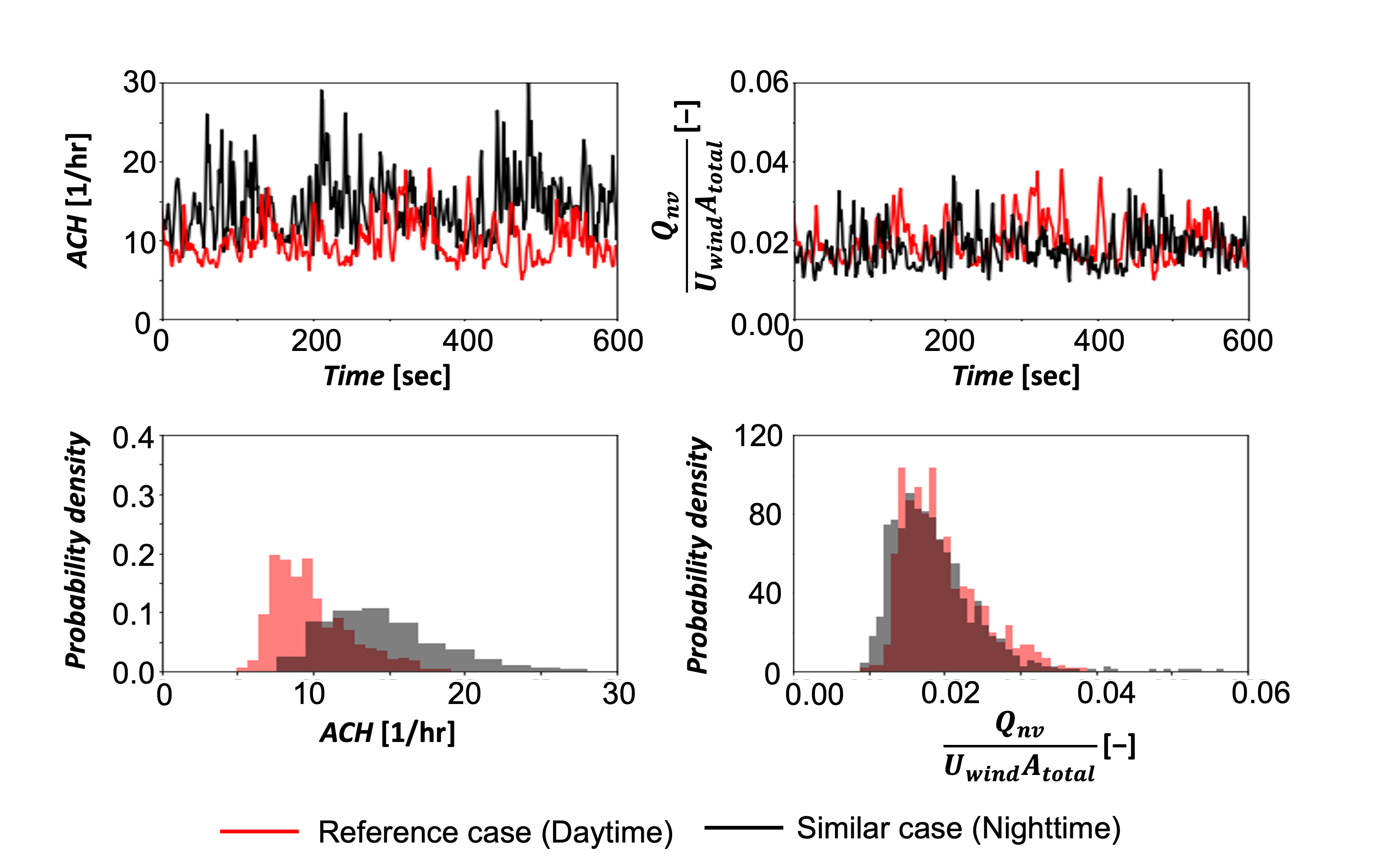}
    \caption{Time series (top) and its frequency distribution (bottom) of ACH (left) and non-dimensional ventilation rate (right) for daytime verification case}
    \label{fig:verification_day}
    ~
    \includegraphics[width=0.9\textwidth]{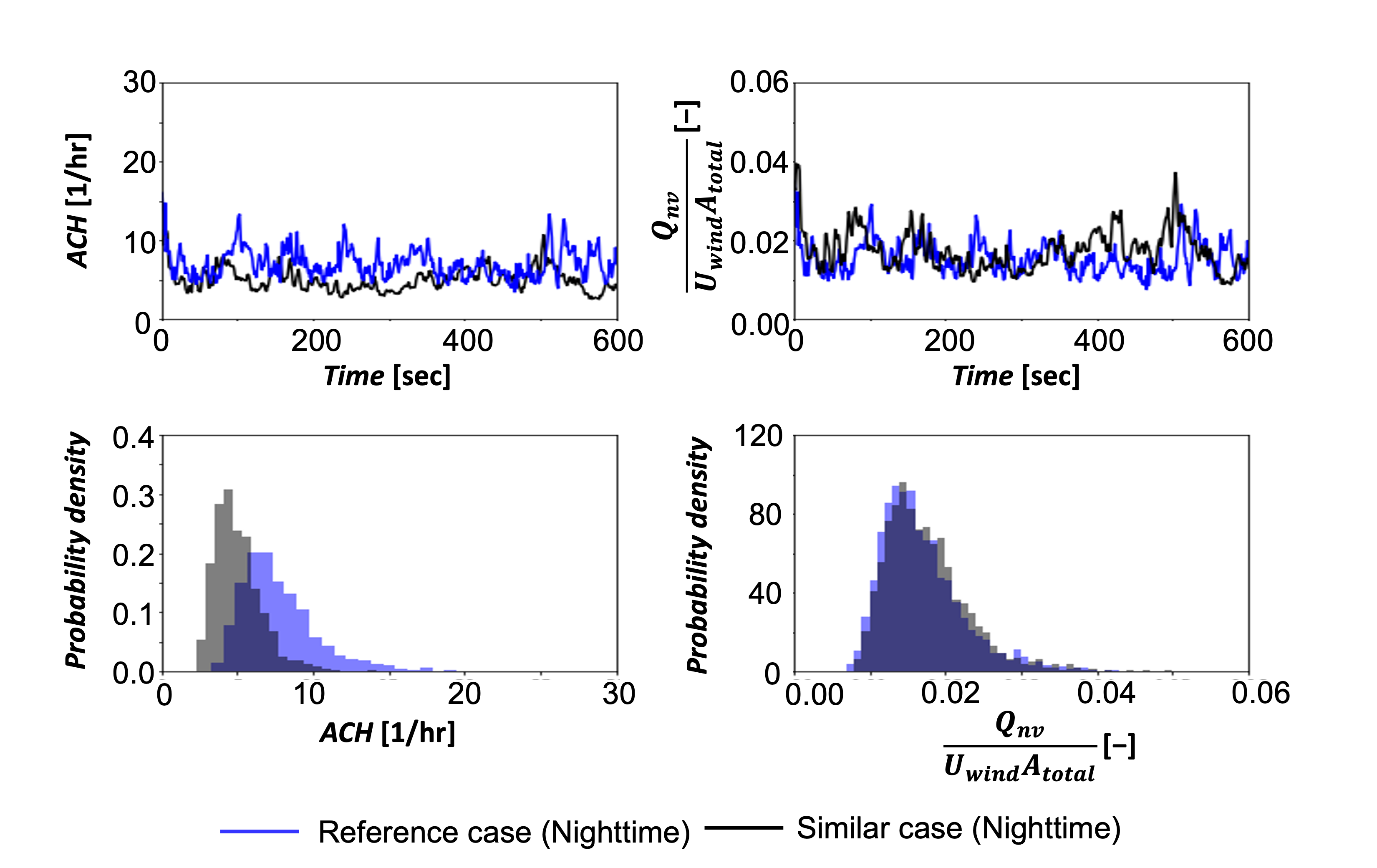}
    \caption{Time series (top) and its frequency distribution (bottom) of ACH (left) and non-dimensional ventilation rate (right) for nighttime verification case}
    \label{fig:verification_night}
\end{figure}

Table~\ref{tab:Ri_similarity_verification} summarizes both the simulation settings and the results for the non-dimensional ventilation rates of the four cases. Figures~\ref{fig:verification_day} and~\ref{fig:verification_night} show the corresponding time series of dimensional and non-dimensional ventilation rates as well as their frequency distribution. The dimensional ACH values differ significantly, but after non-dimensionalizing the time series and distributions collapse. The difference between the mean non-dimensional ventilation rates for the similar cases is 6.5\% for the daytime skylight/window configuration and 3.6\% for the nighttime skylight/floor-level vent configuration. These small differences are likely due to secondary effects of the non-dimensional indoor surface temperature boundary conditions on the ventilation pattern and resulting non-dimensional flow rate; these effects are not represented in the proposed similarity relationship. However, the cost benefit of significantly reducing the parameter space warrants introducing this relatively small (<10\%) uncertainty in the results. 

\subsection{Results: Predictive simulations} \label{subsec:similarity_predictions}
In this section, the functional form of the similarity relationship is established and tested. First, the value of $Ri_v$ is varied with $\theta_{wind}$ fixed to the dominant wind direction; second, the combined effect of $Ri_v$ and $\theta_{wind}$ is determined; third, the ventilation rates obtained from the similarity relationship are compared to the available field measurements. 

\subsubsection{Effect of $Ri_v$ with fixed $\theta_{wind}$} \label{subsubsec:predictions_fixed_wd}
The impact of $Ri_v$ on the non-dimensional ventilation rate $Q'_{nv}$ is investigated by performing simulations for a fixed wind direction of 330$^\circ$, which was the dominant wind direction in the slum neighborhood during the field campaign. 
For the nighttime conditions, we perform simulations with $Ri_v$=[-0.85, -0.6, -0.4, -0.2, -0.0], while for the daytime, we consider $Ri_v$=[+0.0, +0.2, +0.4, +0.6]. The difference between the nighttime and daytime cases is that the temperature boundary conditions are defined differently, using the respective correlations obtained from the measurement data presented in Figure~\ref{fig:temperature_correlation}.

\begin{figure}[htp!]
    \centering
    \includegraphics[width=0.8\textwidth]{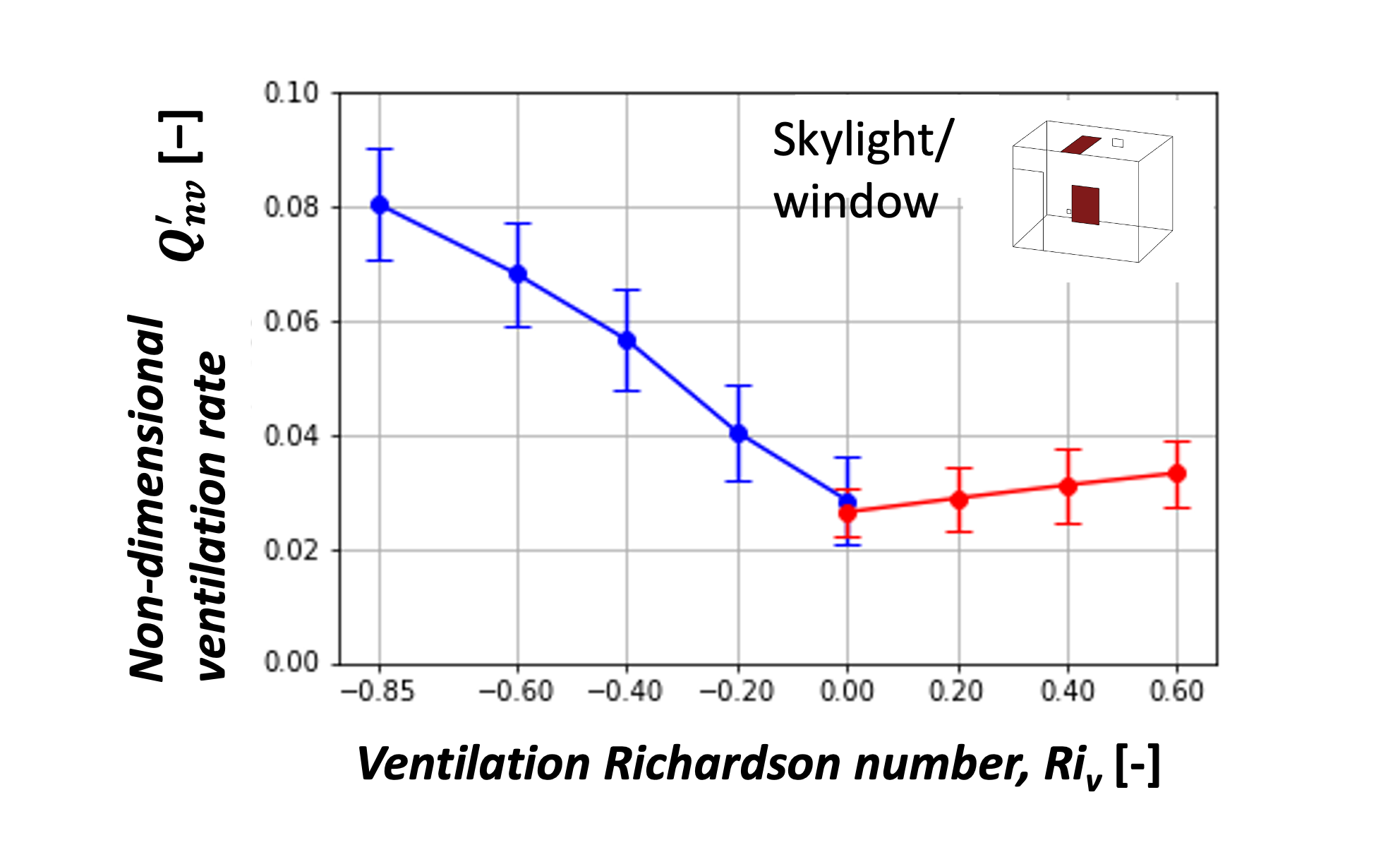}
    \caption{Non-dimensional ventilation rate with respect to ventilation Richardson number with fixed wind direction of 330$^\circ$}
    \label{fig:Ri_results_fixed_WD}
\end{figure}

Figure~\ref{fig:Ri_results_fixed_WD} presents the results, plotting $Q'_{nv}$ as a function of $Ri_v$ for $\theta_{wind}=330^\circ$. The plot represents the mean and standard deviation of the non-dimensional ventilation rate time series predicted by the LES. The two different colors correspond to the daytime cases with indoor thermal stratification (red) and to the nighttime cases with a uniform indoor temperature (blue). The case with $Ri_v = 0.0$ is simulated for both daytime and nighttime conditions, and the small difference between these results confirms that changes in the floor and ceiling temperatures have a limited effect on $Q'_{nv}$. 
Overall, the ventilation rate is lower during daytime than nighttime, primarily due to the different temperature distributions. During the day, the indoor environment is stably stratified, and the neutral line, where the indoor and outdoor temperatures are equal, is close to the roof. This limits buoyancy-driven ventilation, which is reflected in the relatively slow increase of $Q'_{nv}$ with $Ri_v$. During the night, the more neutral stratification of the indoor environment in combination with the negative temperature difference $\Delta T$, supports buoyancy-driven ventilation, resulting in more significant increases in $Q'_{nv}$ as $Ri_v$ becomes more negative.


Although the daytime and nighttime scenarios show a different slope, both cases exhibit roughly linear increases in $Q'_{nv}$ as the absolute value of $Ri_v$ increases over the range of $Ri_v$ considered. It is reasonable to assume that a similar linear dependency will hold for the other wind directions; hence, the following section will consider different wind directions, but only simulating two points for each scenario: $Ri_v$= -0.85, -0.00 for nighttime, and $Ri_v$= +0.00, 0.60 for daytime. 


\subsubsection{Combined effect of $Ri_v$ and $\theta_{wind}$} \label{subsubsec:predictions_ri_wd}

\begin{figure}[htp!]
\centering
    \begin{subfigure}{0.48\textwidth}
    \includegraphics[width=\textwidth]{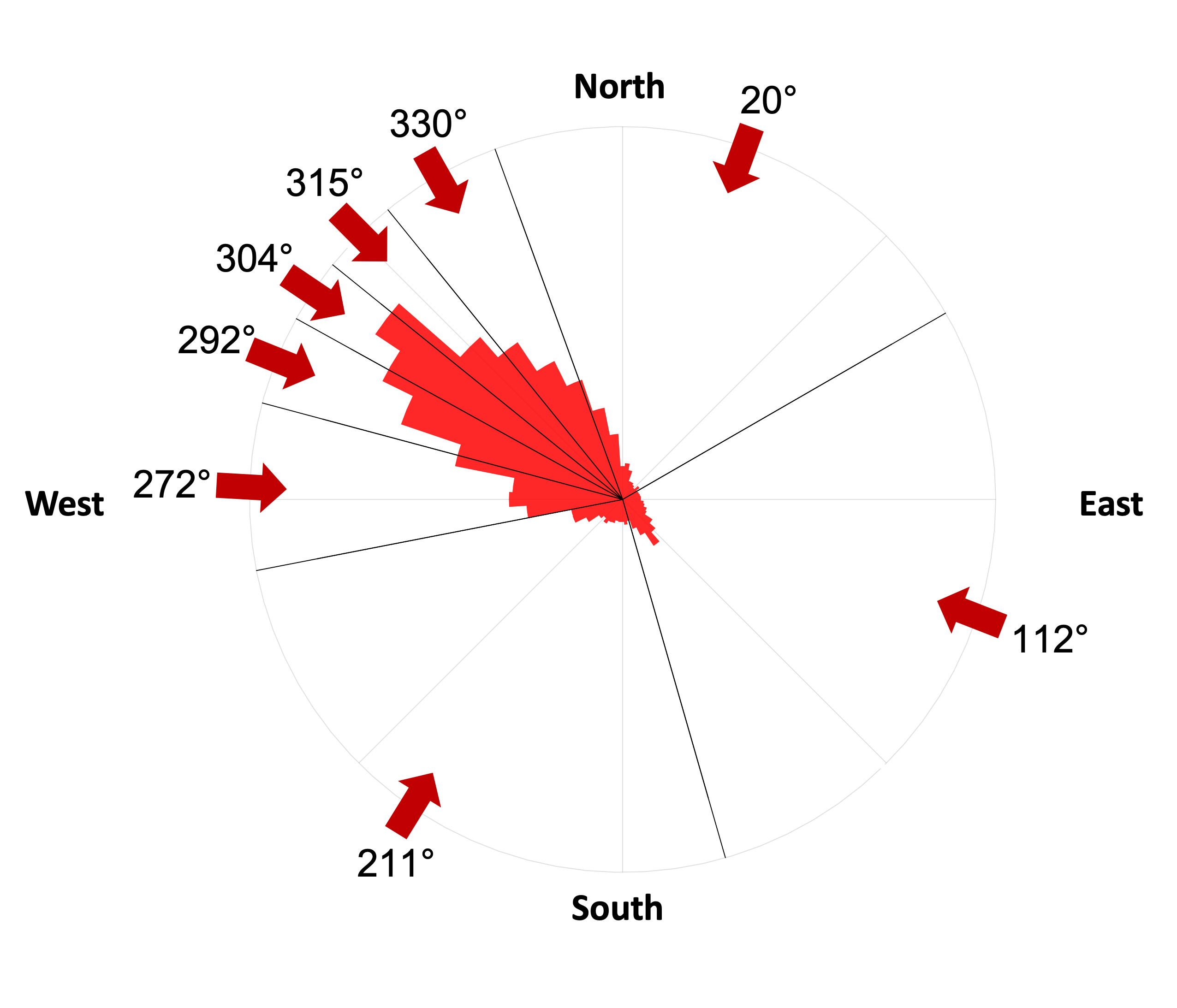}
    \caption{ }
    \label{fig:WD_polar_histogram}
    \end{subfigure}
    ~
    \begin{subfigure}{0.48\textwidth}
    \includegraphics[width=\textwidth]{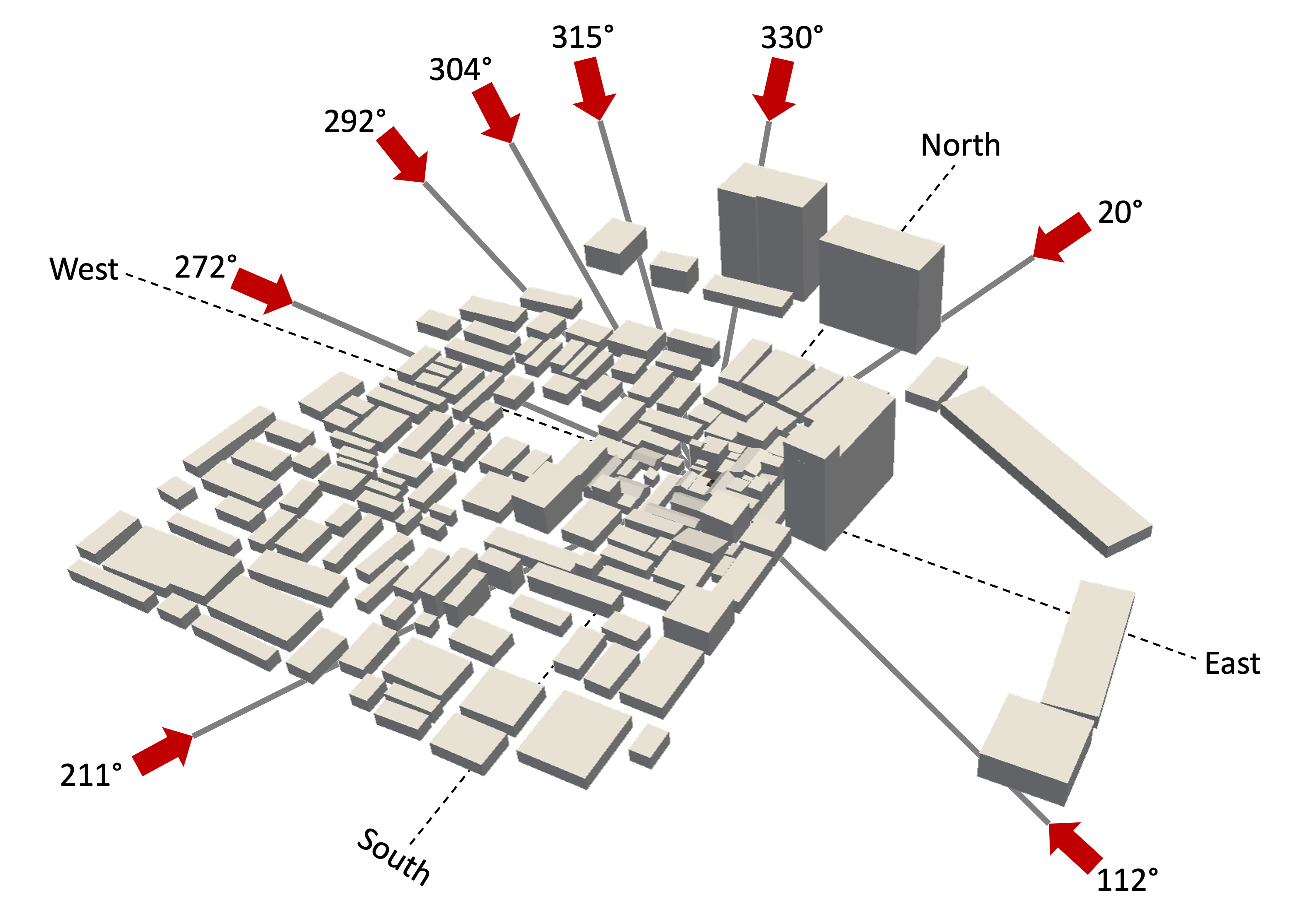}
    \caption{ }
    \label{fig:WD_buildings}
    \end{subfigure}
    \caption{(a) polar-histogram of wind direction data to determine the 8 wind directions to simulate; (b) perspective view of the neighborhood building with the selected wind directions indicated.}
    \label{fig:similarity_WD}
\end{figure}

For wind engineering applications, it is standard practice to explore all wind directions with a 10$^\circ$ resolution. In the context of assessing natural ventilation, there is an opportunity to reduce the number of required simulations by considering the prevalence of the different wind directions at the location of interest. Figure~\ref{fig:WD_polar_histogram} shows the polar histogram of the wind direction data, indicating that the wind is predominantly coming from the north-west. This polar histogram was divided into 8 sectors with an equal probability of occurrence, resulting in a minimum resolution of about 10$^\circ$ around the most dominant wind direction. The median wind directions of these sectors were selected to perform the simulations with the 4 different values for $Ri_v$. This process supports accurate estimates of the natural ventilation flow rates for the most prevalent wind directions, while accepting an increased uncertainty in the estimates for wind directions that rarely occur.  

\begin{figure}[htb!]
    \centering
    \includegraphics[width=\textwidth]{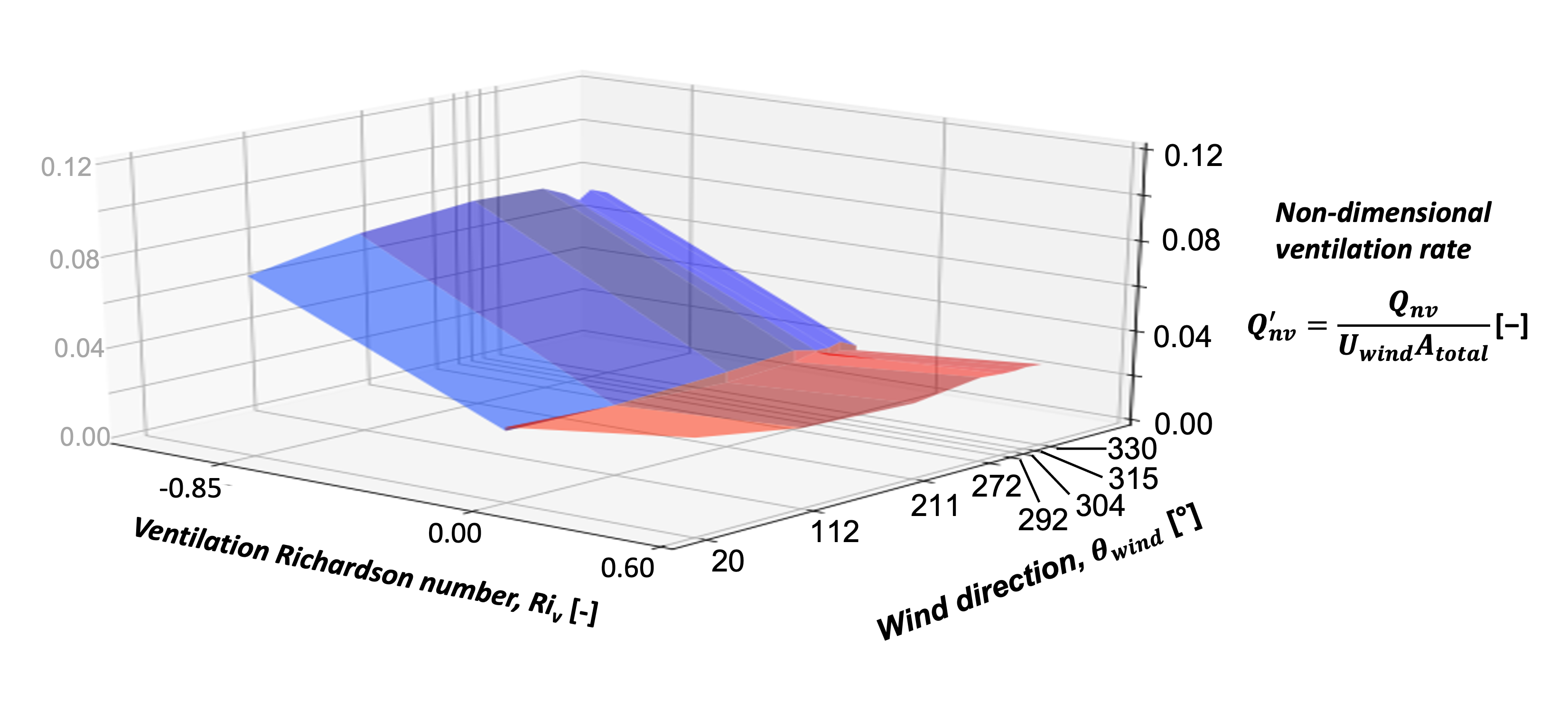}
    \vspace{-30pt}
    \caption{Surrogate model for non-dimensional ventilation rate with respect to $Ri_v$ and $\theta_{wind}$}
    \label{fig:surrogate_Ri_WD}
\end{figure}
The non-dimensional ventilation rates $Q'_{nv}$ obtained from the 32 LES simulations with varying $\theta_{wind}$ and $Ri_v$ are plotted in Figure~\ref{fig:surrogate_Ri_WD}. A striking observation is the small influence of $\theta_{wind}$ compared to the influence $Ri_v$. The maximum variation in $Q'_{nv}$ due to a $\theta_{wind}$ change, observed for $Ri_v=-0.85$, is only about 8\%. In contrast, the increase in $Q'_{nv}$ from $Ri_v=0.00$ to $Ri_v=-0.85$, averaged over all wind directions, is 270\%.

\begin{figure}[bh!] 
    \centering
    \includegraphics[width=1.0\textwidth]{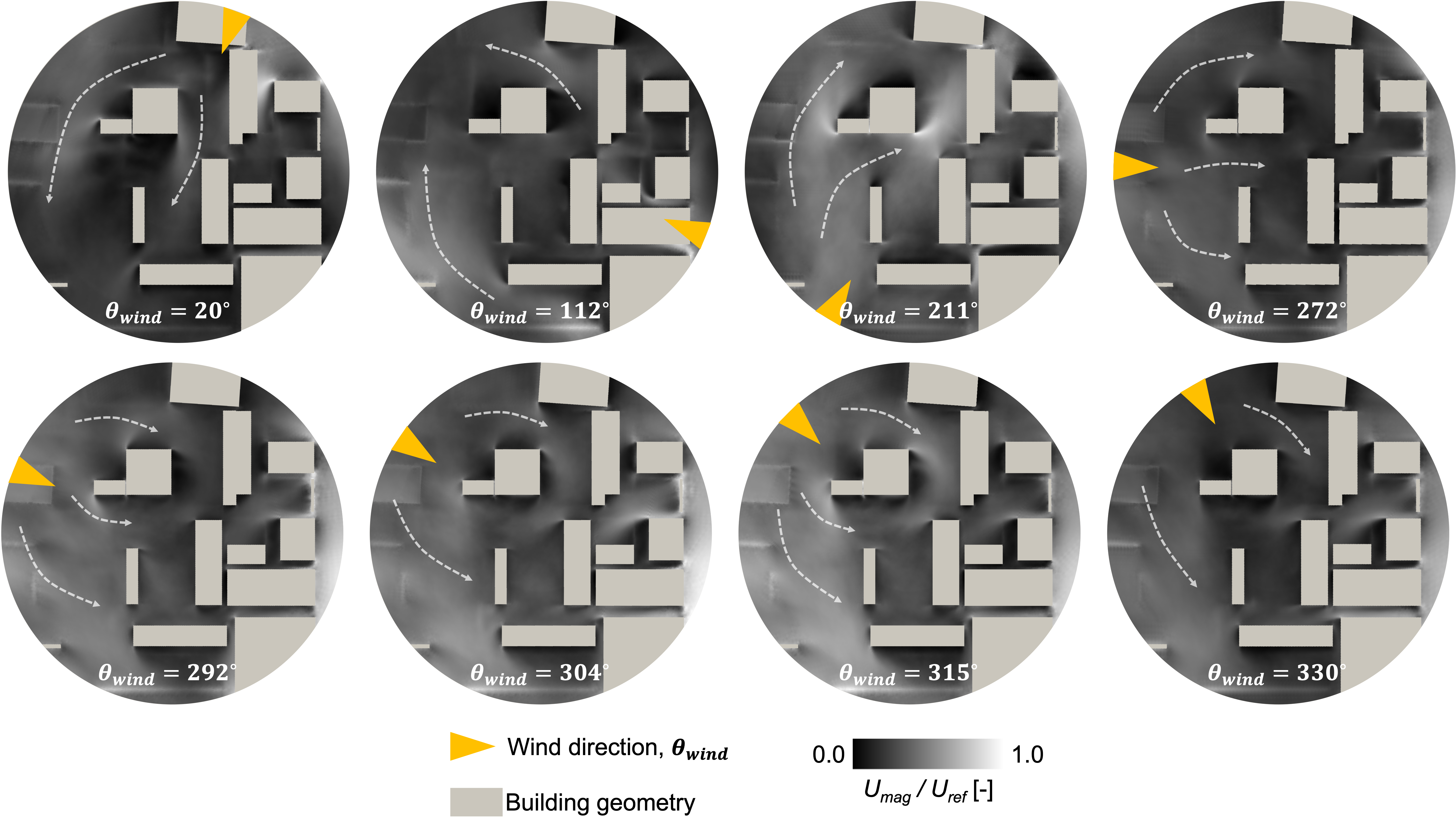}
    \caption{Time-averaged velocity magnitude contours at H=3m within a radius of 15 m from the test house; $Ri_v=-0.00$}
    \label{fig:H3}
\end{figure}

\begin{figure}[thb!] 
    \includegraphics[width=1.0\textwidth]{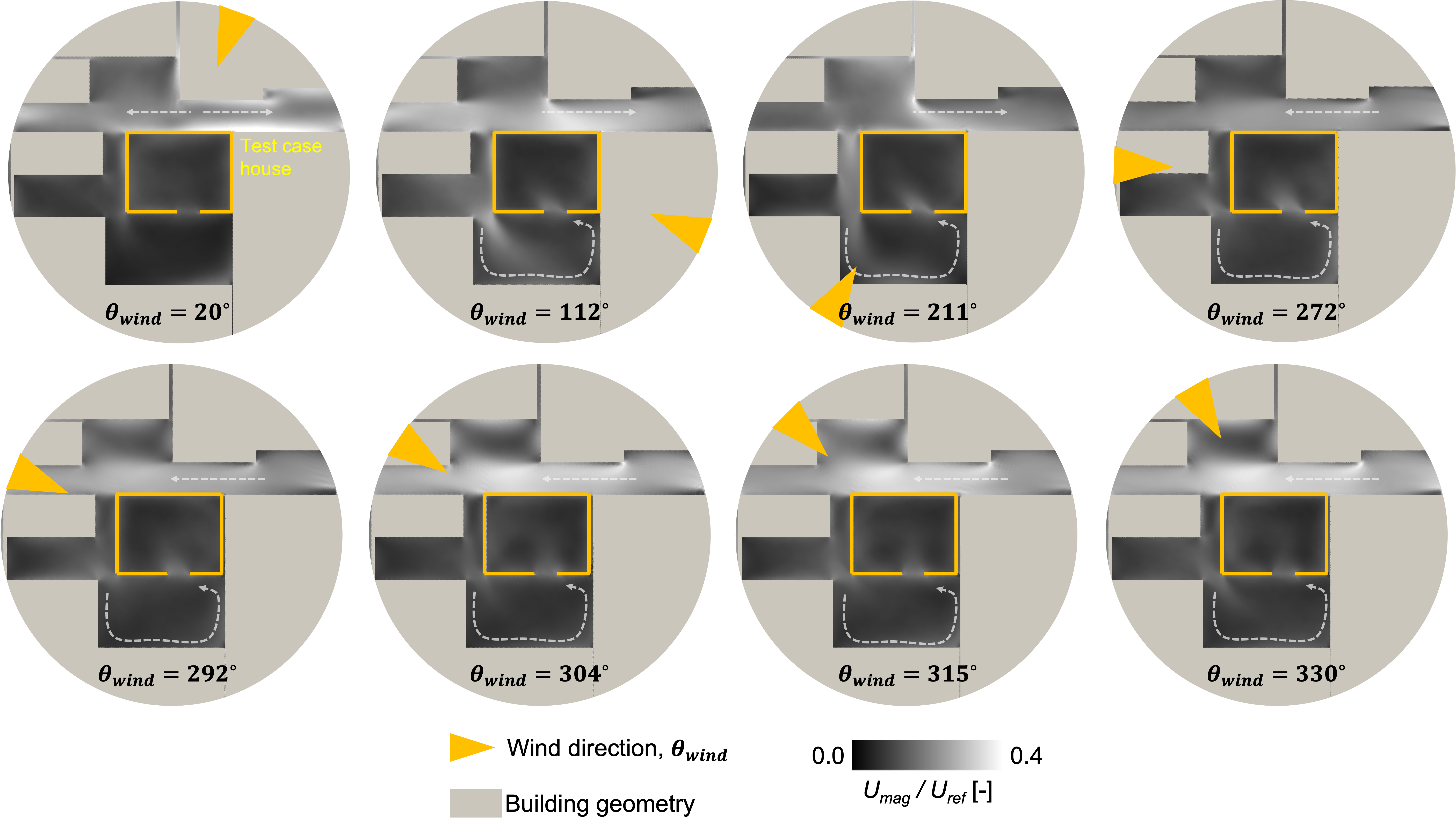}
    \caption{Time-averaged velocity magnitude contours at H=1m within radius of 5 m from the test house; $Ri_v=-0.00$}
    \label{fig:H1}
\end{figure}

The unexpectedly small impact of the wind direction can be tied back to the flow pattern around the test house. Figures~\ref{fig:H3} and~\ref{fig:H1} present the time-averaged velocity fields within a 15 m radius from the test house for all 8 wind directions and $Ri_v=-0.00$ at a height of 3 m and 1 m. The velocity field at 3m height, which is slightly above the average building height, is significantly affected by the incoming wind direction. Buildings that exceed this height are more sparsely distributed, such that the stagnation regions and wake patterns around these buildings change significantly with varying wind directions. However, the flow field at 1m height, which is below the test house roof height, is not significantly different when the wind direction changes. This is especially pronounced for the flow in the courtyard where the window opening is located. These results indicate that in densely packed urban areas, the airflow between buildings is mostly determined by the layout of the urban canopy, such that the impact of the wind direction can be significantly reduced.
 

\subsection{Validation of the similarity relationship}
\begin{figure}[htbp]
\centering
    \includegraphics[width=1.0\textwidth]{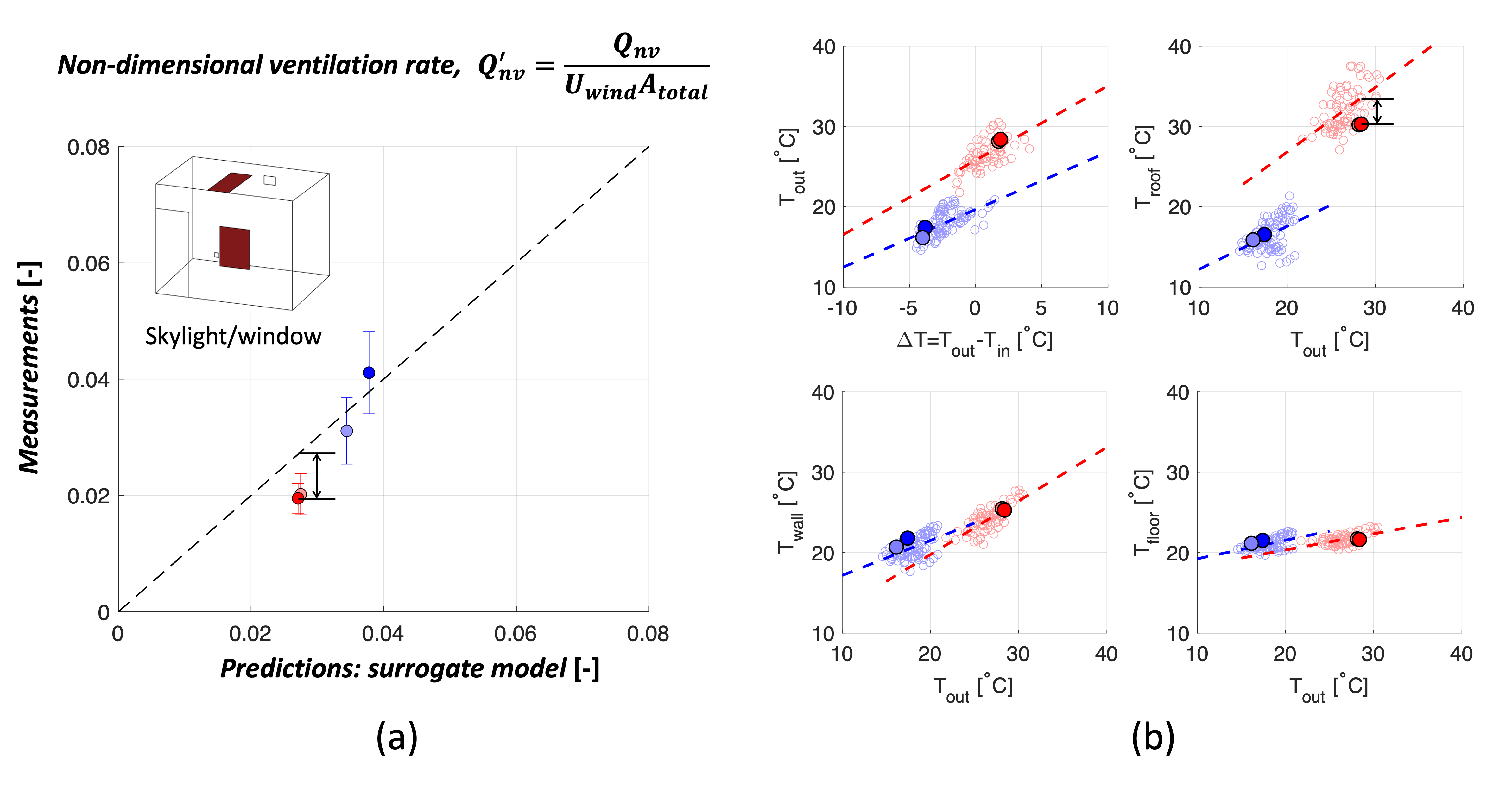}
    \caption{(a) Validation of the surrogate model using $Ri_v$ similarity for the skylight/window configuration; (b) correlation between the different temperature measurements, highlighting the values during the skylight/window ventilation experiments. Daytime cases are shown in red, nighttime cases are shown in blue.}
    \label{fig:similarity_validation_SW}
\end{figure}

The accuracy of the similarity relationship developed in section~\ref{subsubsec:predictions_ri_wd} is evaluated in Figure~\ref{fig:similarity_validation_SW}(a), considering the 4 measurements in the skylight/window configuration. The figure shows a scatter plot of the ACH values obtained by the surrogate model, evaluated at the $Ri_v$ and $\theta_{wind}$ observed during the measurements, vs the measured ACH values. For the nighttime cases (blue/skyblue), the discrepancies between the predicted ACH value and the mean value from the measurement is less than 9\%. Figure~\ref{fig:similarity_validation_SW}(b) shows that during these measurements, the surface temperatures corresponded closely to the values obtained from the correlations used to define the boundary conditions in the LES. For the daytime cases, the discrepancies are slightly higher, with the surrogate model over predicting the measurements by 37\%. This discrepancy is higher than expected based on the validation presented in Section~\ref{sec:LES_validation}. Figure~\ref{fig:similarity_validation_SW}(b) shows that a likely explanation is that the roof temperature recorded during these measurements was about 3.26$^\circ$ lower than the temperature obtained from the correlation used to define the roof temperatures in the LES that informed the surrogate model. As such, the indoor environment was less stratified during these experiments than what would be expected on average based on the correlations, which reduces the ventilation rate. The surrogate model can be expected to be more accurate when considering the average ventilation rate that the home will see over a longer time period, since the correlations represent the average condition.

\begin{figure}[htb]
\centering
    \includegraphics[width=0.8\textwidth]{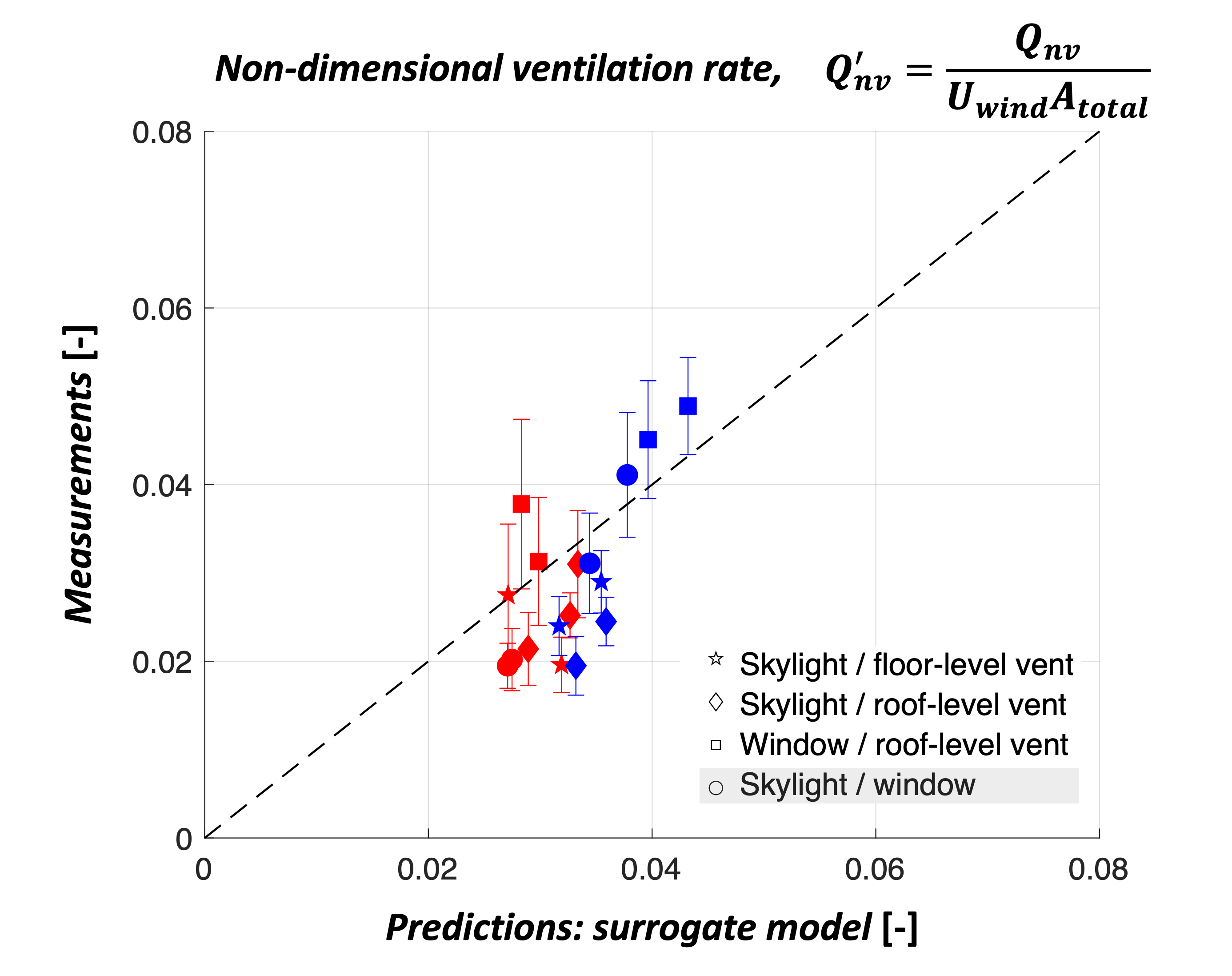}
    \caption{Validation of the surrogate model using $Ri_v$ similarity for all ventilation configurations}
    \label{fig:similarity_validation_all}
\end{figure}
In a final step, we explore whether the surrogate model can predict ventilation in the home more generally, considering all 4 opening configurations for which experiments were performed (see Figure~\ref{fig:outfall_house_drawing}. The aim of this analysis is to provide initial insight into the importance of the specific opening locations in the home. Figure~\ref{fig:similarity_validation_all} presents the comparison of the surrogate model predictions against all 17 measurements under the 4 different configurations. The model correctly captures trend in the measurements, with an average discrepancy of 27\%. This result indicates that both the opening locations, and any variations in the surface temperatures are secondary effects, i.e. the proposed $Ri_v$ similarity relationship captures the dominant physics driving the natural ventilation flow in the test house. 

\section{Conclusion and future work}
This paper has proposed and validated an efficient strategy for using CFD to predict natural ventilation flow rates in a specific house as a function of highly variable boundary conditions. First, urban-scale large-eddy simulations were shown to predict field measurements of the ventilation rate in a single-room urban slum home with a window and a skylight within 25\%. Next, 32 simulations were used to establish a similarity relationship that expresses the dimensionless ventilation rate as a function of just two parameters: the dimensionless ventilation Richardson number and the wind direction. The main assumption in this similarity relationship, namely that for a constant ventilation Richardson number small changes in the non-dimensional indoor temperature field will have a limited effect (<10\%) on the dimensionless ventilation rate, was verified to be correct. The resulting surrogate model indicates a strong dependency of the non-dimensional ventilation rate on the ventilation Richardson number, while the wind direction only has a small secondary effect. The limited effect of the wind direction is attributed to the density of the urban canopy below the roof height of the test building. Comparison of the surrogate model predictions to 4 different field measurements in the configuration with the window and the skylight open reveals differences ranging from 9\% to 37\%. The higher discrepancies occur when the roof surface temperature is lower than average during the field measurements, while the similarity relationship was designed to represent the average conditions. 


In summary, this paper has shown that large-eddy simulations can efficiently inform accurate, building-specific similarity relationships for natural ventilation flow rates. In future work, we will explore the use of a building thermal model to (1) define the correlations for the surface temperature boundary conditions in the absence of field measurements, and (2) obtain predictions of the seasonal and yearly distributions of the ventilation rate. Furthermore, we will investigate how the density of the urban canopy, as well as the specific building configuration and location change the influence of the wind direction on the ventilation rate. The proposed modeling strategy has the potential to support performance-based design of natural ventilation systems to improve occupant health and well-being.

\section*{Acknowledgements}
This research was funded by a seed grant from the Stanford Woods Institute Environmental Venture Projects program and supported by the Stanford Center at the Incheon Global Campus (SCIGC) funded by the Ministry of Trade, Industry, and Energy of the Republic of Korea and managed by the Incheon Free Economic Zone Authority.

\clearpage
\bibliographystyle{unsrt}
\bibliography{reference}

\begin{thebibliography}{10}

\bibitem{wang2016global}
Haidong Wang, Mohsen Naghavi, Christine Allen, Ryan~M Barber, Zulfiqar~A
  Bhutta, Austin Carter, Daniel~C Casey, Fiona~J Charlson, Alan~Zian Chen,
  Matthew~M Coates, et~al.
\newblock Global, regional, and national life expectancy, all-cause mortality,
  and cause-specific mortality for 249 causes of death, 1980--2015: a
  systematic analysis for the global burden of disease study 2015.
\newblock {\em The lancet}, 388(10053):1459--1544, 2016.

\bibitem{ram2014household}
Pavani~K Ram, Dhiman Dutt, Benjamin~J Silk, Saumil Doshi, Carole~B Rudra,
  Jaynal Abedin, Doli Goswami, Alicia~M Fry, W~Abdullah Brooks, Stephen~P Luby,
  et~al.
\newblock {Household air quality risk factors associated with childhood
  pneumonia in urban Dhaka, Bangladesh}.
\newblock {\em The American journal of tropical medicine and hygiene},
  90(5):968--975, 2014.

\bibitem{wang2017assessment}
Jihong Wang, Shugang Wang, Tengfei Zhang, and Francine Battaglia.
\newblock Assessment of single-sided natural ventilation driven by buoyancy
  forces through variable window configurations.
\newblock {\em Energy and buildings}, 139:762--779, 2017.

\bibitem{etheridge2011natural}
David Etheridge.
\newblock {\em Natural ventilation of buildings: theory, measurement and
  design}.
\newblock John Wiley \& Sons, 2011.

\bibitem{karava2004wind}
Panagiota Karava, Ted Stathopoulos, and Andreas~K Athienitis.
\newblock Wind driven flow through openings--a review of discharge
  coefficients.
\newblock {\em International journal of ventilation}, 3(3):255--266, 2004.

\bibitem{seifert2006cross}
Joachim Seifert, Yuguo Li, James Axley, and Markus R{\"{o}}sler.
\newblock {Calculation of wind-driven cross ventilation in buildings with large
  openings}.
\newblock {\em Journal of Wind Engineering and Industrial Aerodynamics},
  94(12):925--947, 2006.

\bibitem{karava2007wind}
Panagiota Karava, Ted Stathopoulos, and Andreas~K Athienitis.
\newblock Wind-induced natural ventilation analysis.
\newblock {\em Solar Energy}, 81(1):20--30, 2007.

\bibitem{caciolo2011full}
Marcello Caciolo, Pascal Stabat, and Dominique Marchio.
\newblock Full scale experimental study of single-sided ventilation: Analysis
  of stack and wind effects.
\newblock {\em Energy and Buildings}, 43(7):1765--1773, 2011.

\bibitem{karava2011airflow}
Panagiota Karava, Ted Stathopoulos, and Andreas~K Athienitis.
\newblock {Airflow assessment in cross-ventilated buildings with operable
  fa{\c{c}}ade elements}.
\newblock {\em Building and Environment}, 46(1):266--279, 2011.

\bibitem{larsen2018calculation}
Tine~Steen Larsen, Christoffer Plesner, Val{\'e}rie Leprince,
  Fran{\c{c}}ois~R{\'e}mi Carri{\'e}, and Anne~Kirkegaard Bejder.
\newblock Calculation methods for single-sided natural ventilation: Now and
  ahead.
\newblock {\em Energy and Buildings}, 177:279--289, 2018.

\bibitem{hu2008cfd}
Cheng-Hu Hu, Masaaki Ohba, and Ryuichiro Yoshie.
\newblock {CFD modelling of unsteady cross ventilation flows using LES}.
\newblock {\em Journal of Wind Engineering and Industrial Aerodynamics},
  96(10-11):1692--1706, 2008.

\bibitem{ramponi2012cfd}
Rubina Ramponi and Bert Blocken.
\newblock {CFD simulation of cross-ventilation for a generic isolated building:
  Impact of computational parameters}.
\newblock {\em Building and Environment}, 53:34--48, 2012.

\bibitem{tominaga2016wind}
Yoshihide Tominaga and Bert Blocken.
\newblock {Wind tunnel analysis of flow and dispersion in cross-ventilated
  isolated buildings: Impact of opening positions}.
\newblock {\em Journal of Wind Engineering and Industrial Aerodynamics},
  155:74--88, 2016.

\bibitem{van2017accuracy}
Twan van Hooff, Bert Blocken, and Yoshihide Tominaga.
\newblock {On the accuracy of CFD simulations of cross-ventilation flows for a
  generic isolated building: comparison of RANS, LES and experiments}.
\newblock {\em Building and Environment}, 114:148--165, 2017.

\bibitem{jiang2002effect}
Yi~Jiang and Qingyan Chen.
\newblock Effect of fluctuating wind direction on cross natural ventilation in
  buildings from large eddy simulation.
\newblock {\em Building and Environment}, 37(4):379--386, 2002.

\bibitem{larsen2011characterization}
Tine~Steen Larsen, Nikos Nikolopoulos, Aris Nikolopoulos, George Strotos, and
  Konstantinos Stefanos~P. Nikas.
\newblock Characterization and prediction of the volume flow rate aerating a
  cross ventilated building by means of experimental techniques and numerical
  approaches.
\newblock {\em Energy and Buildings}, 43(6):1371--1381, 2011.

\bibitem{srebric2008bcs}
Jelena Srebric, Vladimir Vukovic, Guoqing He, and Xudong Yang.
\newblock {CFD boundary conditions for contaminant dispersion, heat transfer
  and airflow simulations around human occupants in indoor environments}.
\newblock {\em Building and Environment}, 43(3):294--303, 2008.

\bibitem{ji2007numerical}
Yingchun Ji and Malcolm~J Cook.
\newblock Numerical studies of displacement natural ventilation in multi-storey
  buildings connected to an atrium.
\newblock {\em Building Services Engineering Research and Technology},
  28(3):207--222, 2007.

\bibitem{chen2009ventilationperformance}
Qingyan Chen.
\newblock {Ventilation performance prediction for buildings: A method overview
  and recent applications}.
\newblock {\em Building and Environment}, 44(4):848--858, apr 2009.

\bibitem{wykes2020effect}
Megan~Davies Wykes, El~Khansaa Chahour, and Paul~F. Linden.
\newblock The effect of an indoor-outdoor temperature difference on transient
  cross-ventilation.
\newblock {\em Building and Environment}, 168:106447, 2020.

\bibitem{xing2001study}
Huijuan Xing, Andy Hatton, and Hazim~B Awbi.
\newblock A study of the air quality in the breathing zone in a room with
  displacement ventilation.
\newblock {\em Building and environment}, 36(7):809--820, 2001.

\bibitem{jiang2003buoyancy}
Yi~Jiang and Qingyan Chen.
\newblock Buoyancy-driven single-sided natural ventilation in buildings with
  large openings.
\newblock {\em International Journal of Heat and Mass Transfer},
  46(6):973--988, 2003.

\bibitem{bangalee2013computational}
Md~Zavid~Iqbal Bangalee, Jiun-Jih Miau, and San-Yih Lin.
\newblock Computational techniques and a numerical study of a buoyancy-driven
  ventilation system.
\newblock {\em International Journal of Heat and Mass Transfer}, 65:572--583,
  2013.

\bibitem{caciolo2012numerical}
Marcello Caciolo, Pascal Stabat, and Dominique Marchio.
\newblock {Numerical simulation of single-sided ventilation using RANS and LES
  and comparison with full-scale experiments}.
\newblock {\em Building and Environment}, 50:202--213, 2012.

\bibitem{Linden1999}
Paul Linden.
\newblock The fluid mechanics of natural ventilation.
\newblock {\em Annual Review of Fluid Mechanics}, 31:201–238, 1999.

\bibitem{etheridge2015}
David Etheridge.
\newblock A perspective on fifty years of natural ventilation research.
\newblock {\em Building and Environment}, 91:51--60, 2015.

\bibitem{islam2006slums}
Nazrul Islam, G~Angeles, A~Mahbub, P~Lance, and NI~Nazem.
\newblock Slums of urban bangladesh: mapping and census 2005.
\newblock 2006.

\bibitem{charles}
{Cascade Technologies, Inc.}
\newblock {CharLES}.
\newblock https://www.cascadetechnologies.com, 2020.

\bibitem{vreman2004eddy}
AW~Vreman.
\newblock An eddy-viscosity subgrid-scale model for turbulent shear flow:
  Algebraic theory and applications.
\newblock {\em Physics of fluids}, 16(10):3670--3681, 2004.

\bibitem{franke2011cost}
Jorg Franke, Antti Hellsten, K~Heinke Schlunzen, and Bertrand Carissimo.
\newblock {The COST 732 Best Practice Guideline for CFD simulation of flows in
  the urban environment: a summary}.
\newblock {\em International Journal of Environment and Pollution},
  44(1-4):419--427, 2011.

\bibitem{tominaga2008aij}
Yoshihide Tominaga, Akashi Mochida, Ryuichiro Yoshie, Hiroto Kataoka, Tsuyoshi
  Nozu, Masaru Yoshikawa, and Taichi Shirasawa.
\newblock {AIJ guidelines for practical applications of CFD to pedestrian wind
  environment around buildings}.
\newblock {\em Journal of wind engineering and industrial aerodynamics},
  96(10-11):1749--1761, 2008.

\bibitem{xie2008efficient}
Zheng-Tong Xie and Ian~P Castro.
\newblock Efficient generation of inflow conditions for large eddy simulation
  of street-scale flows.
\newblock {\em Flow, turbulence and combustion}, 81(3):449--470, 2008.

\bibitem{kim2013divergence}
Yusik Kim, Ian~P Castro, and Zheng-Tong Xie.
\newblock Divergence-free turbulence inflow conditions for large-eddy
  simulations with incompressible flow solvers.
\newblock {\em Computers \& Fluids}, 84:56--68, 2013.

\bibitem{lamberti2018optimizing}
Giacomo Lamberti, Clara Garc{\'\i}a-S{\'a}nchez, Jorge Sousa, and Catherine
  Gorl{\'e}.
\newblock Optimizing turbulent inflow conditions for large-eddy simulations of
  the atmospheric boundary layer.
\newblock {\em Journal of Wind Engineering and Industrial Aerodynamics},
  177:32--44, 2018.

\bibitem{wieringa1992updating}
Jon Wieringa.
\newblock Updating the davenport roughness classification.
\newblock {\em Journal of Wind Engineering and Industrial Aerodynamics},
  41(1-3):357--368, 1992.

\bibitem{stull2012abl}
Roland~B Stull.
\newblock {\em An introduction to boundary layer meteorology}, volume~13.
\newblock Springer Science \& Business Media, 2012.

\bibitem{emes2018estimating}
Matthew Emes, Azadeh Jafari, and Maziar Arjomandi.
\newblock Estimating the turbulence length scales from cross-correlation
  measurements in the atmospheric surface layer.
\newblock In {\em 21st Australasian Fluid Mechanics Conference}, 2018.

\bibitem{jiang2001study}
Yi~Jiang and Qingyan Chen.
\newblock Study of natural ventilation in buildings by large eddy simulation.
\newblock {\em Journal of Wind engineering and industrial aerodynamics},
  89(13):1155--1178, 2001.

\bibitem{boulard2002bcs}
Thierry Boulard, Constantinos Kittas, Jean~Claude Roy, and Shaojin Wang.
\newblock Convective and ventilation transfers in greenhouses, part 2:
  determination of the distributed greenhouse climate.
\newblock {\em Biosystems Engineering}, 83(2):129--147, 2002.

\bibitem{van2010effect}
Twan van Hooff and Bert Blocken.
\newblock On the effect of wind direction and urban surroundings on natural
  ventilation of a large semi-enclosed stadium.
\newblock {\em Computers \& Fluids}, 39(7):1146--1155, 2010.

\bibitem{liu2019cfd}
Sumei Liu, Wuxuan Pan, Qing Cao, Zhengwei Long, Yi~Jiang, and Qingyan Chen.
\newblock {CFD simulations of natural cross ventilation through an apartment
  with modified hourly wind information from a meteorological station}.
\newblock {\em Energy and Buildings}, 195:16--25, 2019.

\bibitem{warren1978ventilation}
Peter~R. Warren.
\newblock Ventilation through openings on wall only, energy conservation in
  housing.
\newblock {\em Coolong, and Ventilating Buildings}, 1:189--206, 1978.

\bibitem{de1982ventilation}
W~De~Gids and H~Phaff.
\newblock Ventilation rates and energy consumption due to open windows: a brief
  overview of research in the netherlands.
\newblock {\em Air infiltration review}, 4(1):4--5, 1982.

\bibitem{warren1984singlesided}
Peter~R. Warren and Lynn~M. Parkins.
\newblock Single-sided ventilation through open windows.
\newblock In {\em Document - Swedish Council for Building Research}, 1984.

\bibitem{hunt1999fluid}
Gary~R. Hunt and Paul~F. Linden.
\newblock The fluid mechanics of natural ventilation—displacement ventilation
  by buoyancy-driven flows assisted by wind.
\newblock {\em Building and Environment}, 34(6):707--720, 1999.

\bibitem{larsen2008single}
Tine~S Larsen and Per Heiselberg.
\newblock Single-sided natural ventilation driven by wind pressure and
  temperature difference.
\newblock {\em Energy and Buildings}, 40(6):1031--1040, 2008.

\bibitem{cen2017revised}
European~Committee for Standardization~(CEN).
\newblock {16798-7:2017 Energy performance of buildings – Ventilation for
  buildings – Part 7: Calculation methods for the determination of air flow
  rates in buildings including infiltration (Modules M5-5)}.
\newblock {\em CEN/TC 156, Brussels, Belgium}, 2017.

\end{thebibliography}

\end{document}